\documentclass[prc,aps,twocolumn,showpacs,amssymb,superscriptaddress,fleqn]{revtex4-1}

\usepackage{amsmath}
\usepackage{color}
\usepackage[table]{xcolor}
\usepackage{graphicx}
\usepackage{dcolumn}
\usepackage{mathtools}
\usepackage{relsize}
\usepackage{float}
\usepackage{braket}

\newcommand{\lwk}{{{\rm low}\mbox{-}k}}
\newcommand{\vlwk}{$V_{{\rm low}\mbox{-}k}$}
\newcommand{\thetaeff}{$\Theta_{\rm eff}$}

\newcommand{\zbb}{$0\nu\beta\beta$}
\newcommand{\dbb}{$2\nu\beta\beta$}
\newcommand{\heff}{$H_{\rm eff}$}

\newcommand{\qbox}{$\hat{Q}$~box}
\newcommand{\tbox}{$\hat{\Theta}$~box}
\newcommand{\nme}{$M^{0\nu}$}
\newcommand{\nmes}{$M^{0\nu}$s}
\newcommand{\nmed}{$M^{2\nu}$}
\newcommand{\nmeds}{$M^{2\nu}$s}

\begin{document}

\title{Shell-model calculation of $^{100}$Mo double-$\beta$ decay}

\author{L. Coraggio}
\affiliation{Dipartimento di Matematica e Fisica, Universit\`a degli
  Studi della Campania ``Luigi Vanvitelli'', viale Abramo Lincoln 5 -
  I-81100 Caserta, Italy}
\affiliation{Istituto Nazionale di Fisica Nucleare, \\ 
Complesso Universitario di Monte  S. Angelo, Via Cintia - I-80126 Napoli, Italy}
\author{N. Itaco}
\affiliation{Dipartimento di Matematica e Fisica, Universit\`a degli
  Studi della Campania ``Luigi Vanvitelli'', viale Abramo Lincoln 5 -
  I-81100 Caserta, Italy}
\affiliation{Istituto Nazionale di Fisica Nucleare, \\ 
Complesso Universitario di Monte  S. Angelo, Via Cintia - I-80126 Napoli, Italy}
\author{G. De Gregorio}
\affiliation{Dipartimento di Matematica e Fisica, Universit\`a degli
  Studi della Campania ``Luigi Vanvitelli'', viale Abramo Lincoln 5 -
  I-81100 Caserta, Italy}
\affiliation{Istituto Nazionale di Fisica Nucleare, \\ 
Complesso Universitario di Monte  S. Angelo, Via Cintia - I-80126 Napoli, Italy}
\author{A. Gargano}
\affiliation{Istituto Nazionale di Fisica Nucleare, \\
Complesso Universitario di Monte  S. Angelo, Via Cintia - I-80126 Napoli, Italy}
\author{R. Mancino}
\affiliation{Dipartimento di Matematica e Fisica, Universit\`a degli
  Studi della Campania ``Luigi Vanvitelli'', viale Abramo Lincoln 5 -
  I-81100 Caserta, Italy}
\affiliation{Istituto Nazionale di Fisica Nucleare, \\ 
Complesso Universitario di Monte  S. Angelo, Via Cintia - I-80126 Napoli, Italy}
\author{F. Nowacki}
\affiliation{Universit\'e de Strasbourg, IPHC, 23 rue du Loess 67037 Strasbourg, France}
\affiliation{CNRS, IPHC UMR 7178, 67037 Strasbourg, France}

\begin{abstract}
For the first time, the calculation of the nuclear matrix element of
the double-$\beta$ decay of $^{100}$Mo, with and without the emission
of two neutrinos, is performed in the framework of the nuclear shell
model.
This task is accomplished starting from a realistic nucleon-nucleon
potential, then the effective shell-model Hamiltonian and decay
operators are derived within the many-body perturbation theory.
The exotic features which characterize the structure of Mo isotopes --
such as shape coexistence and triaxiality softness -- push the
shell-model computational problem beyond its present limits, making it
necessary to truncate the model space. 
This has been done with the goal to preserve as much as possible the
role of the rejected degrees of freedom in an effective approach that
has been introduced and tested in previous studies.
This procedure is grounded on the analysis of the effective
single-particle energies of a large-scale shell-model Hamiltonian,
that leads to a truncation of the number of the orbitals belonging to
the model space.
Then, the original Hamiltonian generates a new one by way of a unitary
transformation onto the reduced model space, to retain effectively the
role of the excluded single-particle orbitals.
The predictivity of our calculation of the nuclear matrix element for
the neutrinoless double-$\beta$ decay of $^{100}$Mo is supported by
the comparison with experiment of the calculated spectra,
electromagnetic transition strengths, Gamow-Teller transition
strengths and the two-neutrino double-$\beta$ nuclear matrix
elements.
\end{abstract}

\pacs{21.60.Cs, 21.30.Fe, 27.60.+j, 23.40-s}

\maketitle

\section{Introduction}
\label{intro}
Between the late 1990s and the early 2000s, the observation that solar
and atmospheric neutrinos oscillate \cite{Fukuda98,Ahmad01} has
indicated that these elusive particles have non-zero mass, and
supported the investigations to search for physics beyond the Standard
Model \cite{Falcone01,Mohapatra06}.
This discovery has revived the interest in the study of neutrinoless
double-$\beta$ decay (\zbb), a rare second-order electroweak process
that, if occuring, would provide fundamental knowledge about the
nature of the neutrino.
In fact, such a decay would demonstrate that neutrinos are Majorana
particles, namely they are their own antiparticles, and violate the
conservation of the lepton quantum number.
Moreover, the measurement of the half-life of \zbb~decay would be a
source of knowledge about the absolute scale of neutrino masses and
their hierarchy, normal or inverted \cite{DellOro15}.

The standard mechanism that is considered in a \zbb~decay is the
exchange of a light Majorana neutrino, and in such a framework the
half-life is expressed as
  
\begin{equation}
\left[ T^{0\nu}_{1/2}\right]^{-1} = G^{0\nu} g_A^4 \left| M^{0\nu} \right|^2
\left|\frac{ \langle m_{\nu}\rangle}{m_e}\right|^2~,
\label{halflife}
\end{equation}

\noindent
where $G^{0\nu}$ is the phase-space factor \cite{Kotila12,Kotila13},
\nme~is the nuclear matrix element directly related to the wave
functions of the parent and grand-daughter nuclei, $g_A$ is the axial
coupling constant, $m_e$ is the electron mass, and $\langle m _{\nu}
\rangle = \sum_i (U_{ei})^2 m_i$ is the effective neutrino mass, as
expressed in terms of the neutrino masses $m_i$ and their mixing
matrix elements $U_{ei}$.

The expression in (\ref{halflife}) makes explicit the crucial role of
the physics of nuclear structure, since the calculation of \nme,
which cannot be measured, provides the value of the neutrino effective
mass in terms of the half-life $T^{0\nu}_{1/2}$ and of the nuclear
structure factor $F_N=G^{0\nu} \left| M^{0\nu} \right|^2g_{A}^4$.
The value of \nme~is also important to estimate the half-life an
experiment should measure in order to be sensitive to a particular
value of the neutrino effective mass \cite{Avignone08}, by combining
the nuclear structure factor, the neutrino mixing parameters
\cite{PDG18}, and present limits on $\langle m_{\nu} \rangle$ from
current observations.

It is then highly desirable that the theory could provide reliable
calculations of \nme, namely that all uncertainties and truncations
which characterize the application of a nuclear model are under
control, leading eventually to an estimate of the theoretical error.
This is currently within reach of {\it ab initio} calculations, but at
present this approach has been pursued mainly for light nuclei
\cite{Pastore18,Cirigliano18,Cirigliano19} whereas the best candidates
of experimental interest are located in the region of medium- and
heavy-mass nuclei.
The nuclear matrix element of \zbb~decay of $^{48}$Ca, the
lightest nuclide of experimental interest, has been also calculated
using both an {\it ab initio} approach which combines the in-medium
similarity renormalization group (IMSRG) with the generator coordinate
method \cite{Yao20a}, and the coupled cluster method \cite{Novario21}.
More recently, a calculation of \nmes~ for the \zbb-decay of
$^{48}$Ca, $^{76}$Ge, and $^{82}$Se has been performed in terms of
in-medium similarity renormalization group \cite{Belley21}. 

Presently, the study of nuclei that are the target of on-going
experiments cannot be performed within the {\it ab initio} framework,
and the nuclear structure models which are mostly employed are the
interacting boson model (IBM) \cite{Barea13}, the quasiparticle
random-phase approximation (QRPA) \cite{Terasaki15,Fang18}, energy density
functional methods (EDF)\cite{Rodriguez10,Rodriguez13}, the covariant
density functional theory \cite{Yao15,Song17}, the
generator-goordinate method (GCM) \cite{Jiao17,Jiao18},
and the shell model (SM)
\cite{Senkov13,Holt13d,Senkov14,Neacsu15,Menendez17,Coraggio20b}.

Among several candidates to the detection of \zbb~decay, $^{100}$Mo is
nowadays one of the most interesting one.
As a matter of fact, $^{100}$Mo is characterized by one of the largest
decay energies ($Q_{\beta\beta} = 3034.36 \pm 0.17$ keV) \cite{Wang17}
which largely suppresses the $\gamma$ background, and its natural
abundance of $9.7\%$ makes experiments, which are targeted to this
nuclide, to be arranged with ton-scale detectors.

Experiments that are searching \zbb~decay of $^{100}$Mo are AMoRE
\cite{Alenkov19,Bhang12}, NEMO 3 \cite{NEMO3}, CUPID-Mo
\cite{Armengaud20a, Armengaud21}, and in a future the ton-scale CUPID
(CUORE Upgrade with Particle IDentification) \cite{CUPID19}.

Recently, the CUPID-Mo experiment has posed a new limit on the
half-life of \zbb~decay in $^{100}$Mo of $T^{0\nu}_{1/2} > 1.5 \times
10^{24}$ yr \cite{Armengaud21}.

Despite the encouranging features as a candidate to the detection of
neutrinoless double-$\beta$ decay, the structure of $^{100}$Mo poses
serious difficulties for a microscopic calculation of the
$\beta$-decay properties of this nuclide and consequently of its
\zbb-decay nuclear matrix element.
As a matter of fact, since 1970s there is experimental evidence for a
rotational behavior of neutron-rich Mo isotopes \cite{Cheifetz70}, and
many nuclear structure studies have been carried out to study their transition
from spherical to deformed shapes, as well as to search for shape
coexistence and triaxiality
\cite{vonBrentano04,Cejnar04,Zhang15,Xiang16,Abusara17}.

Collective models are then better endowed for a satisfactory
description of heavy-mass molybdenum  isotopes than microscopic ones,
and there are few calculations of $^{100}$Mo spectroscopic properties
within the nuclear shell model \cite{Johnstone98,Ozen06}.
As a matter of fact, calculation of $\beta$-decay properties of
$^{100}$Mo and estimates of its \zbb-decay nuclear matrix element have
been carried out within the framework of EDF \cite{Vaquero13,Yao15},
IBM \cite{Barea13,Barea15}, and extensively with QRPA and
proton-neutron QRPA (pn-QRPA)
\cite{Tomoda91,Pantis96,Chaturvedi03,Simkovic13,Hyvarinen15}.

In the present work, for the first time, the study of the
double-$\beta$ decay of $^{100}$Mo is approached from the point of
view of the realistic shell model (RSM) \cite{Coraggio09a}, namely the
effective SM Hamiltonian \heff~ and decay operators are consistently
derived starting from a realistic nucleon-nucleon ($NN$) potential
$V^{NN}$.

The outset is the high-precision CD-Bonn $NN$ potential
\cite{Machleidt01b}, whose repulsive high-momentum components are
renormalized using the \vlwk~procedure \cite{Bogner02}.
The low-momentum \vlwk~is amenable to a perturbative expansion of the
shell-model effective Hamiltonian
\cite{Kuo95,Hjorth95,Suzuki95,Coraggio12a} and decay operators
\cite{Ellis77,Coraggio20c}, so that single-particle (SP) energies, two-body
matrix elements of the residual interaction (TBMEs), matrix elements of
effective electromagnetic transitions and GT-decay operators, as well
as two-body matrix elements of the effective \zbb-decay operator are
derived in terms of a microscopic approach, without adjusting SM
parameters to reproduce data.
This approach has been recently employed first to study two-neutrino
double-$\beta$ (\dbb) decay of $^{48}$Ca, $^{76}$Ge, $^{82}$Se,
$^{130}$Te, and $^{136}$Xe \cite{Coraggio17a,Coraggio19a}, and then to
calculate \nmes~of the same nuclides for their \zbb~decay
\cite{Coraggio20a}.

The model space we choose to calculate the nuclear wave functions of
$^{100}$Mo and $^{100}$Ru, which are the main characters of the decay
process we investigate in this work, is spanned by four
$0f_{5/2},1p_{3/2},1p_{1/2},0g_{9/2}$ proton orbitals and five
$0g_{7/2},1d_{5/2},1d_{3/2},2s_{1/2},0h_{11/2}$ neutron orbitals
outside $^{78}$Ni core, which is characterized by the $Z=28,N=50$
shell closures.
This means that the structure of $^{100}$Mo should be described in terms
of 14 and 8 valence protons and neutrons, respectively, interacting in
such a large model space, while $^{100}$Ru is characterized by 16 and
6 valence protons and neutrons.

It has to be noted that such a model space may be not large enough to
account for the ground-state deformation of nuclei around $A \sim 100$
such as $^{100}$Zr \cite{Sieja09}, and that perhaps $Z,N=50$
cross-shell excitations should be explicitly included to reproduce the
large observed $B(E2;2^+_1 \rightarrow 0^+_1)$ values \cite{Coraggio16a}.
However, as we will see in Section \ref{results}, this choice of the
model space does not seem to affect the overall comparison between the
experimental and our calculated $B(E2)$s, both for $^{100}$Mo and
$^{100}$Ru.

The computational problem owns a high degree of difficulty, being at
the limit of actual capabilities and burdensome to handle.
Then, we have employed a procedure that aims to reduce the
computational complexity of large-scale shell-model calculations, by
preserving effectively the role of the rejected degrees of
freedom.
First, the truncation is driven by the analysis of the effective SP
energies (ESPE) of the original Hamiltonian, so to locate the
relevant degrees of freedom to describe $A=100$ Mo,Tc, and Ru
isotopes, namely the single-particle orbitals that will constitute a
smaller and manageable model space.
As a second step, we perform an unitary transformation of the
original Hamiltonian, defined in the model space that is made up
respectively by four and five proton and neutron orbitals (labelled as
$[45]$), onto the truncated model space.
This transformation generates a new shell-model Hamiltonian that, even
if defined within a smaller number of configurations, retains
effectively the role of the excluded SP orbitals.

This double-step procedure, that is to derive a first \heff~in a large
space and then from this a new one in a smaller space, has been
introduced in Refs. \cite{Coraggio15a,Coraggio16a} for nuclei in the
mass region $A \approx 100$ outside $^{88}$Sr core, and successfully
applied also for Mo isotopes up to $A=98$ \cite{Coraggio16a}.

In the following section we outline first the derivation of \heff~and
SM effective decay operators by way of the many-body perturbation
theory.
Then, we sketch out some details about the double-step procedure to
derive a new \heff~in a smaller space, and show an example aimed to support
its validity.
In Section \ref{results} we report the calculated low-energy
spectroscopic properties of the nuclei involved in the double-$\beta$
decay process under investigation, the parent and grand-daughter nuclei
$^{100}$Mo,Ru, as well as the calculated GT-strength distributions and
\nmeds, and compare them with available data.
In the same section we report the results of the calculation of
\nme~for $^{100}$Mo, together with an analysis of the angular
momentum-parity matrix-element distributions, and a comparison with the
results obtained with other nuclear structure models.
Finally, the last section is devoted to a summary of the present work
and an outlook of our future developments.

\section{Theoretical framework}\label{outline}
\subsection{The effective SM Hamiltonian}\label{effham}
The starting point of our calculation is the high-precision CD-Bonn
$NN$ potential \cite{Machleidt01b}, whose repulsive high-momentum
components -- that prevent a perturbative approach to the many-body
problem -- are renormalized by way of the \vlwk~approach
\cite{Bogner02,Coraggio09a}.

This unitary transformation provides a smooth potential that preserves
the values of all $NN$ observables calculated with the CD-Bonn
potential, as well as the contribution of the short-range correlations
(SRC).
The latters account for the action of a two-body decay operator on an
unperturbed (uncorrelated) wave function, which is employed to derive
the SM effective \zbb~ operator, that is different from acting the same
operator on the real (correlated) nuclear wave function.
The details about the treatment of SRC consistently with the
\vlwk~ transformation are reported in
Refs. \cite{Coraggio20a,Coraggio20b,Coraggio20d}.

The \vlwk~matrix elements are then employed as interaction vertices of
the perturbative expansion of \heff, and detailed surveys about this topic
can be found in Refs. \cite{Hjorth95,Coraggio12a,Coraggio20c}.
Here, we sketch briefly the procedure that has been followed to derive
\heff~and SM effective decay operators.

We begin by considering the full nuclear Hamiltonian for $A$ interacting
nucleons $H$, which, within the nuclear shell model, is broken up as a
sum of a one-body term $H_0$, whose eigenvectors set up the SM basis,
and a residual interaction $H_1$, by way of harmonic-oscillator (HO)
one-body potential $U$:

\begin{eqnarray}
 H &= & T + V_\lwk = (T+U)+(V_\lwk-U)= \nonumber \\
~& = &H_{0}+H_{1}~.\label{smham}
\end{eqnarray}

Since this Hamiltonian cannot be diagonalized for a many-body system
in an infinite basis of eigenvectors of $H_0$, we derive an effective
Hamiltonian, which operates in a truncated model space that, in order
to obtain a satisfactory description of $^{100}$Mo, is spanned by four
proton -- $0f_{5/2}, 1p_{3/2}, 1p_{1/2}, 0g_{9/2}$ -- and five neutron
orbitals -- $0g_{7/2}, 1d_{5/2}, 1d_{3/2}, 2s_{1/2}, 0h_{11/2}$ --
outside $^{78}$Ni core.
From now on, we dub this model space as $[45]$.

To this end, we perform a similarity transformation which provides,
within the full Hilbert space of the configurations, a decoupling of
the model space $P$, where the valence nucleons are constrained, from
its complement $Q=1-P$.

This may be obtained within the time-dependent perturbation theory,
namely we derive \heff~through the Kuo-Lee-Ratcliff folded-diagram
expansion in terms of the \qbox~vertex function
\cite{Kuo90,Hjorth95,Coraggio12a}:

\begin{equation}
H^{\rm eff}_1 (\omega) = \hat{Q}(\epsilon_0) - P H_1 Q \frac{1}{\epsilon_0
  - Q H Q} \omega H^{\rm eff}_1 (\omega) ~, \label{eqfinal}
\end{equation}
\noindent
where $\omega$ is the wave operator decoupling the $P$ and $Q$
subspaces, and $\epsilon_0$ is the eigenvalue of the unperturbed
degenerate Hamiltonian $H_0$.

The \qbox~is defined as
\begin{equation}
\hat{Q} (\epsilon) = P H_1 P + P H_1 Q \frac{1}{\epsilon - Q H Q} Q
H_1 P ~, \label{qbox}
\end{equation}
\noindent
and $\epsilon$ is an energy parameter called ``starting energy''.

An exact calculation of the \qbox~is computationally prohibitive, so
the term $1/(\epsilon - Q H Q)$ is expanded as a power series

\begin{equation}
\frac{1}{\epsilon - Q H Q} = \sum_{n=0}^{\infty} \frac{1}{\epsilon -Q
  H_0 Q} \left( \frac{Q H_1 Q}{\epsilon -Q H_0 Q} \right)^{n} ~,
\end{equation}

\noindent
namely we perform an expansion of the \qbox~up to the third order in
perturbation theory \cite{Coraggio20c}.

Then, the \qbox~is the building block to solve the non-linear matrix
equation (\ref{eqfinal}) to derive \heff~ through iterative techniques
such as the Kuo-Krenciglowa and Lee-Suzuki ones
\cite{Krenciglowa74,Suzuki80}, or graphical non-iterative methods
\cite{Suzuki11}.

\begin{figure}[H]
\begin{center}
\includegraphics[scale=0.60,angle=0]{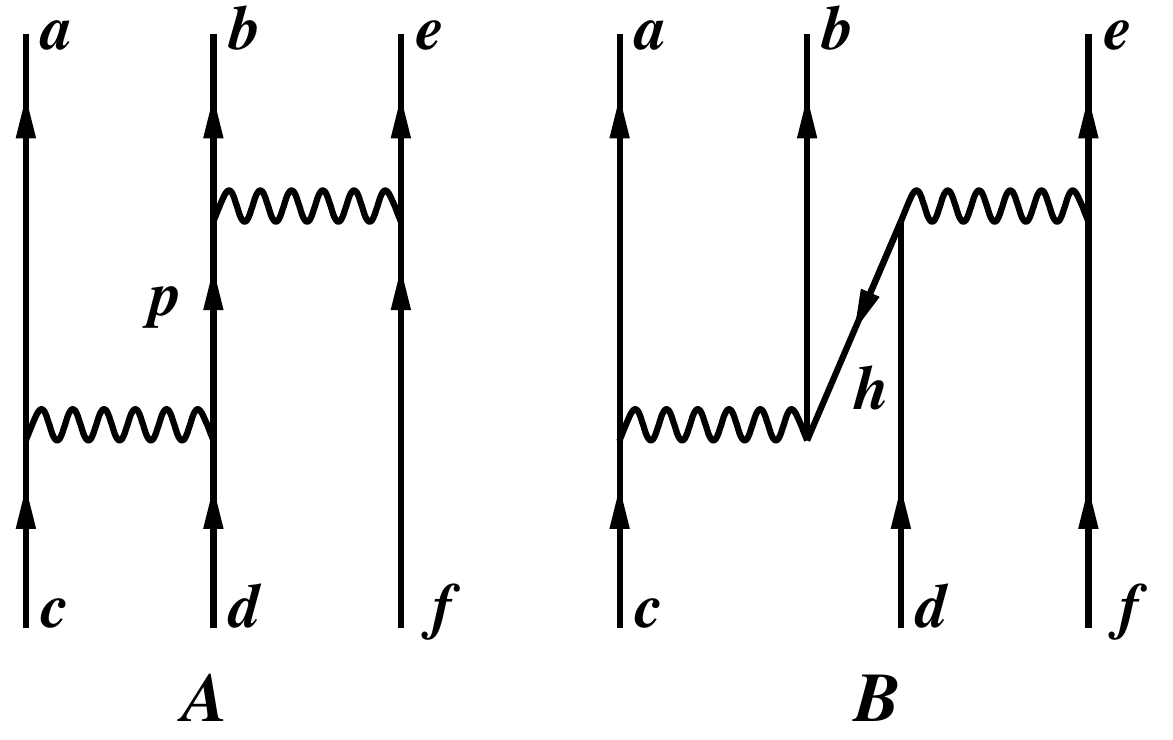}
\caption{Second-order three-body diagrams. The sum over the
  intermediate lines runs over particle and hole states outside the
  model space. For each topology A and B, it has been reported only
  one of the diagrams which correspond to the permutations of the
  external lines.}
\label{diagram3corr}
\end{center}
\end{figure}

This theoretical framework has been well established for systems with
one- and two-valence nucleon systems, but, because of the choice of
the model space, the nuclei that are involved in the decay process
under investigation -- $^{100}$Mo,Tc,Ru -- are characterized by 22
valence nucleons.
Then, one should derive a many-body \heff~which depends on this number
of valence particles, and introduce a formalism that may become very
difficult to be managed.
A minimal choice is to include in the calculation of the \qbox~at
least contributions from three-body diagrams, which account for the
interaction via the two-body force of the valence nucleons with
configurations outside the model space (see Fig. \ref{diagram3corr}).

Since we employ the SM code ANTOINE to calculate the spectra and
double $\beta$-decay nuclear matrix elements \cite{ANTOINE}, a
diagonalization of a three-body \heff~cannot be performed and  we
derive a density-dependent two-body term from the three-body
contribution arising at second order in perturbation theory.
The details of such an approach can be found in
Refs. \cite{Coraggio20c,Coraggio20e}, as well as a discussion about
the role of such contributions to the eigenvalues of the SM
Hamiltonian.

In the Introduction we have pointed out that the current limits of the
available SM codes prevent the calculation of the nuclear matrix
elements of double-$\beta$ decay within the $[45]$ model space.
In order to overcome this computational difficulty, we perform a
truncation of the number of SP orbitals following a method we have
introduced in Ref. \cite{Coraggio15a}, and whose details may be found
in Ref. \cite{Coraggio16a}.

We now sketch the main steps of this procedure.

First, we study the evolution of the proton and/or neutron ESPE as a
function of the valence nucleons, that may justify the exclusion of
one or more SP levels from the original model space (in our case
$[45]$).
Since $^{100}$Mo is described in terms of 14 valence protons and 8
valence neutrons with respect to $^{78}$Ni, this means that a
truncation may be applied only to the number of the neutron orbitals.

\begin{table}[H]
\caption{Theoretical proton and neutron SP energy spacings (in MeV)
  from $H_{\rm eff}^{[45]}$.}
\begin{ruledtabular}
\begin{tabular}{cccc}
Proton orbitals & $\epsilon_p$ & Neutron orbitals & $\epsilon_n$ \\
\colrule
 $0f_{5/2}$ & 0.0  & $0g_{7/2}$  & 2.8 \\ 
 $1p_{3/2}$ & 1.6  & $1d_{5/2}$  & 0.4 \\ 
 $1p_{1/2}$ & 2.1  & $1d_{3/2}$  & 1.1 \\ 
 $0g_{9/2}$ & 4.3  & $2s_{1/2}$  & 0.0 \\
        ~        &   ~   & $0h_{11/2}$ & 3.2 \\
\end{tabular}
\end{ruledtabular}
\label{spetab}
\end{table}

\begin{figure}[H]
\begin{center}
\includegraphics[scale=0.34,angle=0]{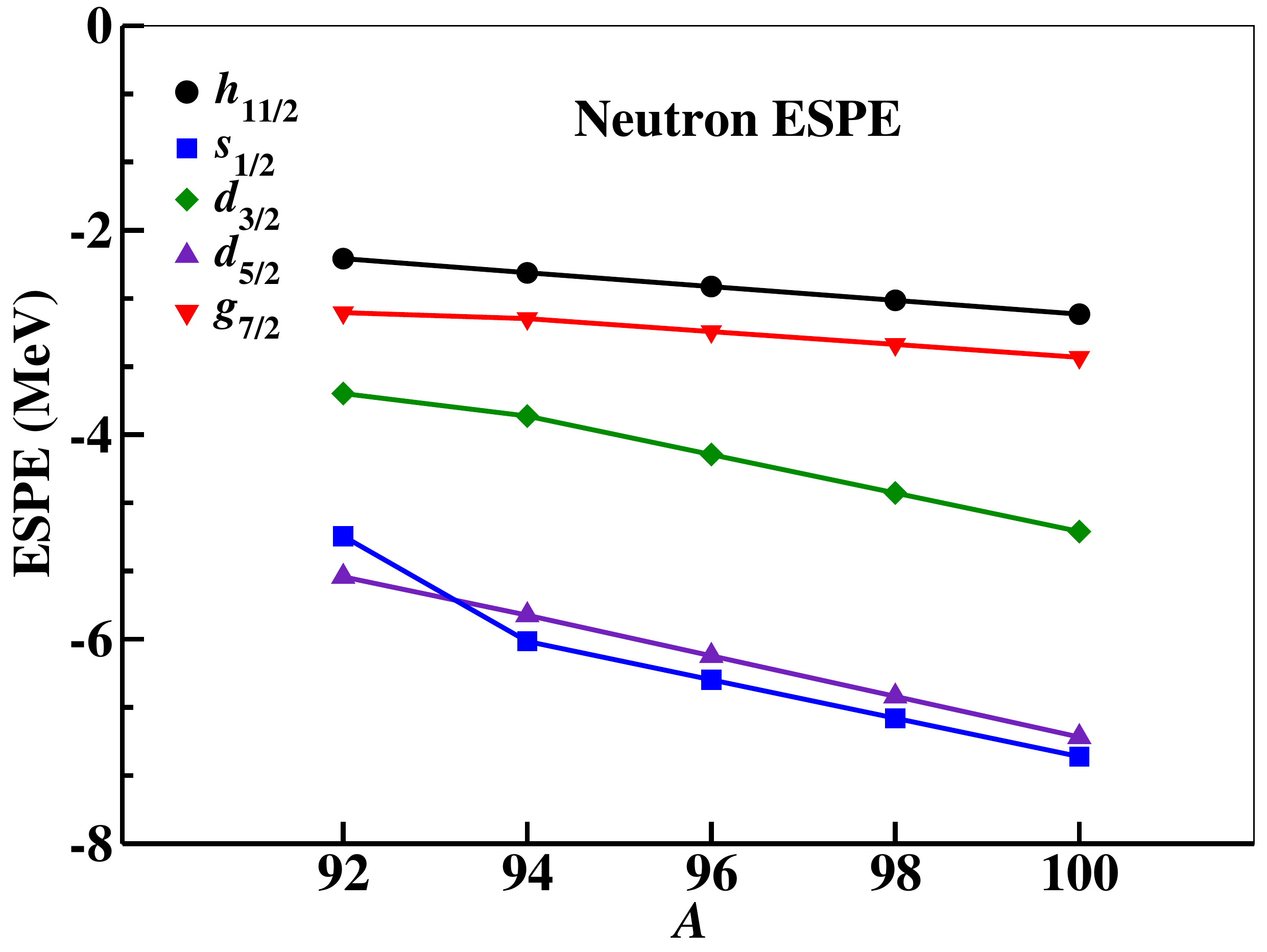}
\caption{Neutron effective single-particle energies of Mo isotopes
  calculated with $H_{\rm eff}^{[45]}$.}
\label{espeMo}
\end{center}
\end{figure}

In Table \ref{spetab} we report the SP energy spacings calculated
using the effective Hamiltonian $H_{\rm eff}^{[45]}$, which is
defined within the model space $[45]$, and in Fig. \ref{espeMo} we
show the behavior of the neutron ESPE of the Mo isotopes.
From the inspection of the table and the figure, we observe that there
is an energy gap separating the $1d_{5/2},1d_{3/2},2s_{1/2}$ neutron
orbitals from the $0g_{7/2},0h_{11/2}$ ones, which enlarges by
increasing the number of valence neutrons.

Therefore, we deem it reasonable the possibility to exclude both
$0g_{7/2}$ and $0h_{11/2}$ neutron orbitals, and deal with a smaller
model space that should still provide the relevant features of the
physics of the nuclei under investigation, namely the parent and
grand-daughter nuclei $^{100}$Mo,Ru.
However, to calculate the nuclear matrix element for the two-neutrino
double-$\beta$ decay \nmed~of $^{100}$Mo we need to retain at least
the neutron $0g_{7/2}$ orbital in the model space, otherwise the
selection rules of the GT operator would forbid such a decay because
of the choice of the proton model subspace.

\begin{figure}[H]
\begin{center}
\includegraphics[scale=0.40,angle=0]{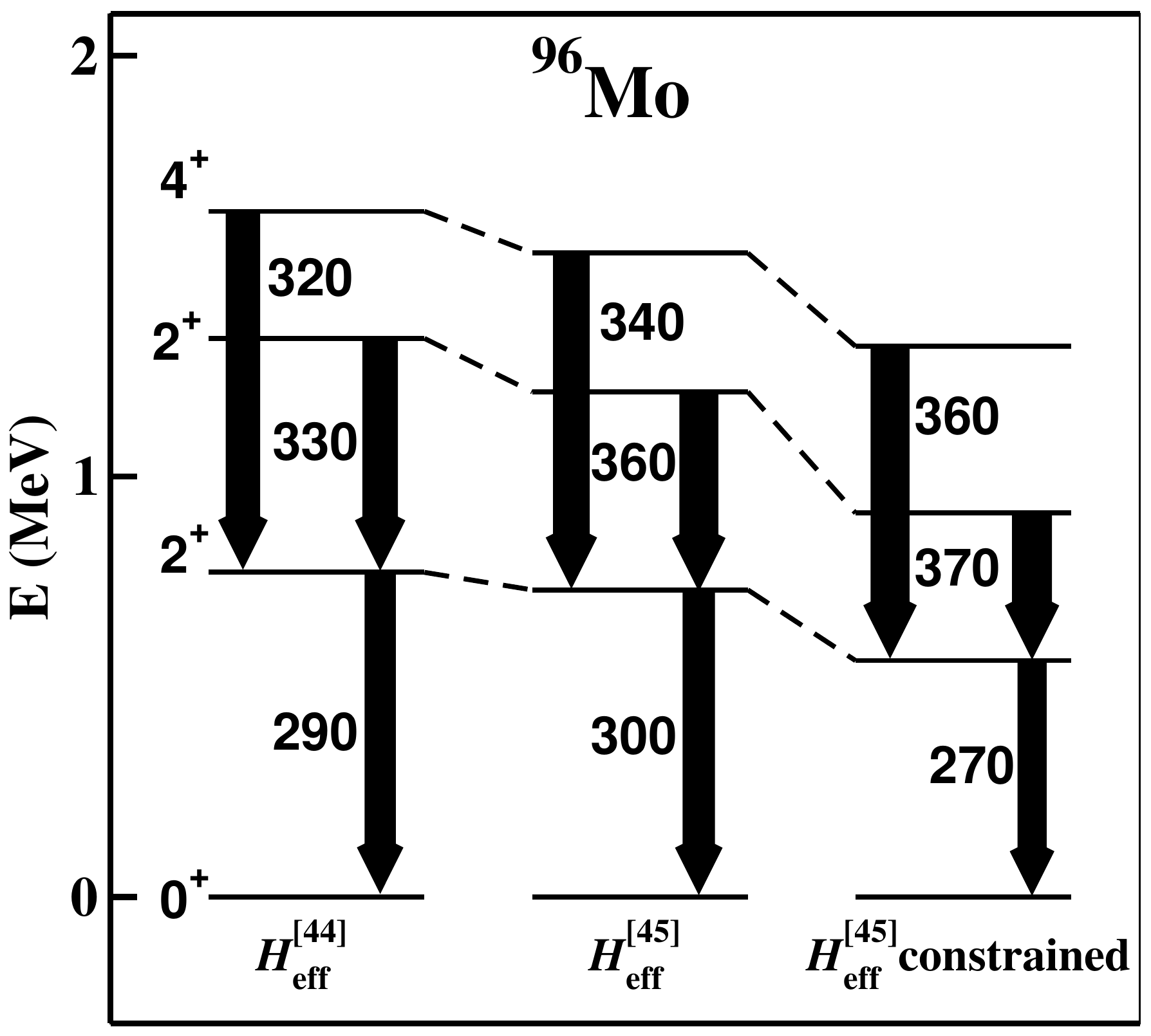}
\caption{Low-energy spectrum of $^{96}$Mo, calculated with $H_{\rm
    eff}^{[45]}$, $H_{\rm eff}^{[44]}$, and constraining $H_{\rm
    eff}^{[45]}$ in the $[44]$ model space. They are reported also the
  value of the significant $B(E2)$ transition rates in $e^2{\rm fm}^4$.}
\label{96Mo}
\end{center}
\end{figure}

On these grounds, we derive a new effective Hamiltonian $H_{\rm
  eff}^{[44]}$, defined within a model space spanned by the
$0f_{5/2},1p_{3/2},1p_{1/2},0g_{9/2}$ proton and
$0g_{7/2},1d_{5/2},1d_{3/2},2s_{1/2}$ neutron orbitals, by way of a
unitary transformation of $H_{\rm eff}^{[45]}$ (see details in
Ref. \cite{Coraggio16a}).
We label this smaller model space $[44]$ and in Fig. \ref{96Mo} we
have reported the energy spectrum of $^{96}$Mo, that is calculated
employing the $H_{\rm eff}^{[45]}$ and $H_{\rm eff}^{[44]}$, and also
constraining the action of $H_{\rm eff}^{[45]}$ in the $[44]$ model
space.

From the inspection of Fig. \ref{96Mo}, it can be noted that $H_{\rm
  eff}^{[44]}$ is able to provide a better agreement with the energy
spectrum obtained through the ``mother Hamiltonian'' $H^{[45]}$ than the
results provided by constraining the diagonalization of the latter
Hamiltonian to model space $[44]$.
It is also worth pointing out that the values of the $B(E2)$
transition rates, that are calculated with $H_{\rm eff}^{[45]}$ and
$H_{\rm eff}^{[44]}$, are very close.

The above results evidence the adequacy of the truncation scheme we
have adopted, and the diagonalization of the SM Hamiltonian for
$^{100}$Mo and $^{100}$Ru has been performed by way of $H_{\rm
  eff}^{[44]}$.

The TBMEs of $H_{\rm eff}^{[44]}$, that have been calculated also
including three-body correlations to account the number of valence
nucleons characterizing $^{100}$Mo, can be found in the Supplemental
Material \cite{supplemental2021}.

\subsection{Effective shell-model decay operators}\label{effopsec}
We are interested not only in calculating energies, but also the
matrix elements of decay operators $\Theta$ which are connected to
measurable quantities such as $B(E2)$ strengths, and the nuclear
matrix element of the \dbb~decay \nmed, as well as the \zbb~decay
matrix element \nme.

Since the diagonalization of the \heff~does not provide the true
wave-functions, but their projections onto the chosen model space $P$,
we need to renormalize any decay operator $\Theta$ to take into
account the neglected degrees of freedom corresponding to the
$Q$-space.

The derivation of SM effective operators within a perturbative
approach dates back to the earliest attempts to employ realistic
potentials for SM calculations
\cite{Mavromatis66,Mavromatis67,Federman69,Ellis77,Towner83,Towner87},
and we follow the procedure that has been introduced by Suzuki and
Okamoto in Ref. \cite{Suzuki95}.
This allows a calculation of decay operators \thetaeff~which is
consistent with the one we carry out of \heff, and that is based on
perturbative expansion of a vertex function \tbox, analogously with
the derivation of \heff~in terms of the \qbox~(see section
\ref{effham}).
The procedure has been reported in details in Ref. \cite{Coraggio20c},
and in the following we only report the main building blocks.

The starting point is the perturbative calculation of the two
energy-dependent vertex functions

\[
\hat{\Theta} (\epsilon) = P \Theta P + P \Theta Q
\frac{1}{\epsilon - Q H Q} Q H_1 P ~, \]
\[ \hat{\Theta} (\epsilon_1 ; \epsilon_2) = P H_1 Q
\frac{1}{\epsilon_1 - Q H Q} Q \Theta Q \frac{1}{\epsilon_2 - Q H Q} Q H_1 P ~,\]

\noindent
and of their derivatives calculated in $\epsilon=\epsilon_0$,
$\epsilon_0$ being the eigenvalue of the degenerate unperturbed
Hamiltonian $H_0$:

\[
\hat{\Theta}_m = \frac {1}{m!} \frac {d^m \hat{\Theta}
 (\epsilon)}{d \epsilon^m} \biggl|_{\epsilon=\epsilon_0} ~, \]
\[ \hat{\Theta}_{mn} =  \frac {1}{m! n!} \frac{d^m}{d \epsilon_1^m}
\frac{d^n}{d \epsilon_2^n}  \hat{\Theta} (\epsilon_1 ;\epsilon_2)
\biggl|_{\epsilon_1= \epsilon_0, \epsilon_2  = \epsilon_0} ~\]

Then, a series of operators $\chi_n$ is calculated:

\begin{eqnarray}
\chi_0 &=& (\hat{\Theta}_0 + h.c.)+ \hat{\Theta}_{00}~~,  \label{chi0} \\
\chi_1 &=& (\hat{\Theta}_1\hat{Q} + h.c.) + (\hat{\Theta}_{01}\hat{Q}
+ h.c.) ~~, \nonumber \\
\chi_2 &=& (\hat{\Theta}_1\hat{Q}_1 \hat{Q}+ h.c.) +
(\hat{\Theta}_{2}\hat{Q}\hat{Q} + h.c.) + \nonumber \\
~ & ~ & (\hat{\Theta}_{02}\hat{Q}\hat{Q} + h.c.)+  \hat{Q}
\hat{\Theta}_{11} \hat{Q}~~, \label{chin} \\
&~~~& \cdots \nonumber
\end{eqnarray}

\noindent
that allows to write \thetaeff~in the following form:
\begin{equation}
\Theta_{\rm eff} = H_{\rm eff} \hat{Q}^{-1}  (\chi_0+ \chi_1 + \chi_2 +\cdots) ~~.
\label{effopexp}
\end{equation}

In this work we arrest the $\chi_n$ series at $n=2$, and the
$\hat{\Theta}$ function is expanded up to third order in perturbation
theory.

\begin{figure}[H]
\begin{center}
\includegraphics[scale=0.40,angle=0]{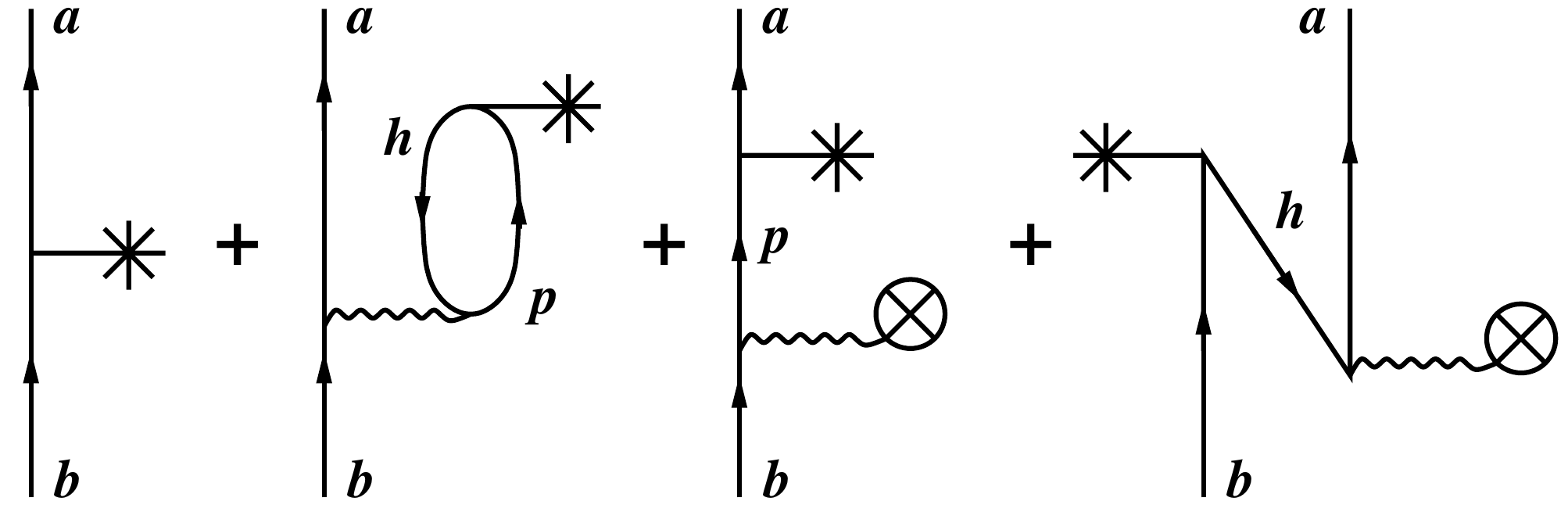}
\caption{One-body second-order diagrams included in the perturbative
  expansion of $\hat{\Theta}(\epsilon_0)$. The asterisk indicates the
  bare operator $\Theta$, the wavy lines the two-body $NN$
  interaction, the circle with a cross inside accounts for the
  ($V\mbox{-}U$)-insertion contribution (see Ref. \cite{Coraggio20c}).}
\label{figeffop1}
\end{center}
\end{figure}

The issue of the convergence of the $\chi_n$ series and of the
perturbative expansion of the \tbox~has been treated in
Refs. \cite{Coraggio18,Coraggio19a,Coraggio20a}, and in
Fig. \ref{figeffop1} they are reported all the diagrams up to second order
appearing in the $\hat{\Theta}(\epsilon_0)$ expansion for a one-body
operator $\Theta$.

In the present work, the decay operators $\Theta$ are the one-body
electric-quadrupole transition $E2$ $q_{\scriptscriptstyle{p,n}}r^2
Y^2_m(\hat{r})$ -- the charge $q_{\scriptscriptstyle {p,n}}$ being $e$
for protons and 0 for neutrons -- and GT $\vec{\sigma} \tau^{\pm}$
operators, as well as the two-body transition operator for the \zbb
decay (see Eqs. (\ref{operatorGT} -- \ref{operatorT}) in the following
subsection).

\subsection{The \dbb- and \zbb-decay operators}\label{operators}
This section is devoted to outline the structure of \dbb-~and
\zbb-decay operators.

It is worth pointing out that these two nuclear-decay mechanisms
differ in the characteristic value of the momentum transfer, which for
\dbb~decay is few MeVs, at variance with the order of hundreds of
MeVs in \zbb~decay.
This difference, as we will see in the following, affects the
procedure to be followed to calculate \nmed~and \nme.

As is well known, \dbb~ decays are the occurrence of two
single-$\beta$ decay transitions inside a nucleus, and the expressions
of the GT and Fermi components of their nuclear matrix elements
\nmed~are the following
\begin{eqnarray}
M_{\rm GT}^{2\nu} & = & \sum_n \frac{ \langle 0^+_f || (\vec{\sigma}
                        \tau^-)_{\scriptscriptstyle{\mathcal I}}
  || 1^+_n \rangle \langle 1^+_n || (\vec{\sigma}
\tau^-)_ {\scriptscriptstyle{\mathcal I}} || 0^+_i \rangle } {E_n + E_0} ~~,\label{doublebetameGT} \\
M_{\rm F}^{2\nu} & = & \sum_n \frac{ \langle 0^+_f || (\tau^-)_{\scriptscriptstyle{\mathcal I}}
  || 0^+_n \rangle \langle 0^+_n || (\tau^-)_{\scriptscriptstyle{\mathcal I}} || 0^+_i \rangle } {E_n +
  E_0} ~~,\label{doublebetameF}
\end{eqnarray}

\noindent
where the subscript $\mathcal I$ indicates we are employing the matrix
elements of either the bare or the effective one-body GT and F operators.

In these equations, $E_n$ is the excitation energy of the $J^{\pi}=0^+_n,1^+_n$ 
intermediate state, and $E_0=\frac{1}{2}Q_{\beta\beta}(0^+) +\Delta M$, where
$Q_{\beta\beta}(0^+)$ and $\Delta M$ are the $Q$ value of the transition
and the mass difference of the parent and daughter nuclear states,
respectively.
The index $n$ runs over all possible intermediate states induced by the given 
transition operator.
It should be pointed out that the Fermi component plays a marginal
role \cite{Haxton84,Elliott02} and in most calculations is neglected
altogether.

The most efficient way to obtain \nmed, by including a number of
intermediate states that is sufficient to provide the needed accuracy
for its calculation, is the Lanczos strength-function method
\cite{Caurier05} which we have adopted for our calculations.

The evaluation of \nmed~could be also carried out employing the
so-called closure approximation, commonly adopted to study \zbb-decay
NMEs \cite{Haxton84}.
On these grounds, within such an approximation the energies of the
intermediate states, $E_n$, appearing in
Eqs.~(\ref{doublebetameGT},\ref{doublebetameF}), may be replaced by an
average value $E_n+E_0 \rightarrow \langle E \rangle$, that allows to
avoid to explicitly calculate the intermediate $J^{\pi}=1^+_n$ states,
but then the two one-body transition operators become a two-body
operator.

Actually, the closure approximation is a valuable tool to evaluate
\nme, since in the \zbb~decay the neutrino's momentum is about one
order of magnitude larger than the average excitation energy of the
intermediate states.
This allows to neglect, within this process, the
intermediate-state-dependent energies from the energy  denominator
appearing in the neutrino potential, as we will see in a while.
On the contrary, the closure approximation is unsatisfactory when used
to calculate  \nmed, because, as mentioned before, the momentum
transfer in \dbb~process is much smaller.

Once the theoretical value on \nmed~has been calculated, it can be
then compared with the experimental counterpart, which is extracted
from the observed half life $T^{2\nu}_{1/2}$ 

\begin{equation}
\left[ T^{2\nu}_{1/2} \right]^{-1} = G^{2\nu} \left| M_{\rm GT}^{2\nu}
\right|^2 ~~,
\label{2nihalflife}
\end{equation}
\noindent
$G^{2\nu}$ being the \dbb-decay phase-space (or kinematic) factor
\cite{Kotila12,Kotila13}.

We now turn our attention to the bare \zbb~operator, for the
light-neutrino-exchange channel~\cite{Engel17}.

The formal expression of $M_{\alpha}^{0\nu}$ -- where $\alpha$ stands
for Fermi ($F$), Gamow-Teller (GT), or tensor ($T$) decay channels --
is written in terms of the one-body transition-density matrix elements
between the daughter and parent nuclei (grand-daughter and daughter
nuclei) $ \langle k | a^{\dagger}_{p^\prime} a_{n^\prime} | i \rangle$ ($ \langle f |
a^{\dagger}_{p}a_{n} | k \rangle $).
The subscripts $p$ and $n$ denote proton and neutron states, and
$i,k,f$ refer to the parent, daughter, and grand-daughter nuclei,
respectively.

The nuclear matrix element $M_{\alpha}^{0\nu}$ is formulated as
\cite{Senkov13,Simkovic08}:
\begin{align}
  M_\alpha^{0\nu} =  
&\sum_{k {\cal J}} \sum_{j_p j_{p^\prime} j_n
    j_{n^\prime} } (-1)^{j_n + j_{n^\prime}+ \cal{J}} \hat{\cal{J}} \left\{
\begin{array}{ccc}
j_p & j_n & J_\kappa \\
j_{n^\prime} & j_{p^\prime} & {\cal J}
\end{array}
                              \right\} \nonumber \\
  ~ & \left< j_p  j_{p^\prime} ;{\cal J} \mid \mid 
\Theta_\alpha^{k} \mid \mid j_n
      j_{n^\prime}; {\cal J} \right>  \nonumber \\
  ~ & \langle k || [a^{\dagger}_{p} \otimes \tilde{a}_{n}]_{J_k} || i
      \rangle \langle k || [a^{\dagger}_{n^\prime} \otimes
      \tilde{a}_{p^\prime}]_{J_k} || f \rangle ^{\ast} = \nonumber \\
~ & \sum_{k} \sum_{j_p j_{p^\prime} j_n  j_{n^\prime} }
   \langle f | a^{\dagger}_{p}a_{n} | k \rangle \langle k |
    a^{\dagger}_{p^\prime} a_{n^\prime} | i \rangle \times \nonumber \\
  ~ & \left< j_p j_{p^\prime} \mid \Theta_\alpha^{k} \mid j_n
      j_{n^\prime} \right> ~, \label{M0nu}
\end{align}

\noindent
where the tilde denotes a time-conjugated state, $\tilde{a}_{j m} =
(-1)^{j+m}a_{j -m}$, and the $\Theta_{\alpha}^k$ are two-body
operators.

The expression of the operators $\Theta_{\alpha}^{k}$ is \cite{Engel17}:
\begin{eqnarray} 
 \Theta^{k}_{\rm GT} & = & [ \tau^-_{1} \tau^-_{2} (\vec{\sigma}_1 \cdot \vec{\sigma}_2) H_{\rm
GT}^k(r) ]_{\scriptscriptstyle{\mathcal I}}\label{operatorGT} \, , \\
\Theta^{k}_{\rm F} & = & [\tau^-_{1} \tau^-_{2} H_{\rm F}^k(r)
                         ]_{\scriptscriptstyle{\mathcal
                         I}} \label{operatorF} \, ,\\
\Theta^{k}_{\rm T} & = & [\tau^-_{1} \tau^-_{2} \left(
                         3\left(\vec{\sigma}_1 \cdot \hat{r} \right)
                         \left(\vec{\sigma}_1 \cdot \hat{r} \right) -
                         \right. \nonumber \\
  ~ & ~ & \left. \vec{\sigma}_1 \cdot \vec{\sigma}_2 \right) H_{\rm
          T}^k(r) ]_{\scriptscriptstyle{\mathcal
          I}}~, \label{operatorT}
\end{eqnarray}

\noindent
where $H_{\alpha}$ are the  neutrino potentials and are defined as:
\begin{equation}
H_{\alpha}^k(r)=\frac {2R}{\pi} \int_{0}^{\infty} \frac {j_{n_{\alpha}}(qr)
  h_{\alpha}(q^2)qdq}{q+E_k-(E_i+E_f)/2}~,
\label{neutpot}
\end{equation}
\noindent
and, again, the subscript ${\mathcal I}$ labels the application of
either the bare or the effective two-body decay operators.

In Eq. (\ref{neutpot}),  $R=1.2 A^{1/3}$ fm, $j_{n_{\alpha}}(qr)$ is
the spherical Bessel function, $n_{\alpha}=0$ for Fermi and
Gamow-Teller components, while $n_{\alpha}=2$ for the tensor
component.
In the following, we also present the explicit expressions of neutrino
form functions, $h_{\alpha}(q)$, for light-neutrino exchange
\cite{Engel17} :
\begin{eqnarray}
h_{\rm F} ({ q}^{2})  & = & g^2_V({ q}^{2}) \, ,  \nonumber \\
h_{\rm GT} ({ q}^{2}) & = & \frac{g^2_A({ q}^{2})}{g^2_A} 
\left[ 1 - \frac{2}{3} \frac{ { q}^{2}}{ { q}^{2} + m^2_\pi } + 
\frac{1}{3} ( \frac{ { q}^{2}}{ { q}^{2} + m^2_\pi } )^2 \right]
\nonumber\\
&& + \frac{2}{3} \frac{g^2_M({ q}^{2} )}{g^2_A} \frac{{ q}^{2} }{4 m^2_p }, 
\nonumber \\
h_{\rm T} ({ q}^{2}) & = & \frac{g^2_A({ q}^{2})}{g^2_A} \left[ 
\frac{2}{3} \frac{ { q}^{2}}{ { q}^{2} + m^2_\pi } -
\frac{1}{3} ( \frac{ { q}^{2}}{ { q}^{2} + m^2_\pi } )^2 \right] 
\nonumber\\
&& + \frac{1}{3} \frac{g^2_M ({ q}^{2} )}{g^2_A} \frac{{ q}^{2} }{4 m^2_p }  \, ,   
\end{eqnarray}
In the present work, we use the dipole approximation for the vector,
$g_V({ q}^{2})$, axial-vector, $g_A({ q}^{2})$, and weak-magnetism,
$g_M({ q}^{2})$, form factors:
\begin{eqnarray} 
g_V({ q}^{2})&=& \frac{g_V}{(1+{ q}^{2}/{\Lambda^2_V})^2}, 
\nonumber\\
g_M({ q}^{2}) &=& (\mu_p-\mu_n) g_V({ q}^{2}), 
\nonumber\\
g_A({ q}^{2}) &=& \frac{g_A}{(1+{ q}^{2}/{\Lambda^2_A})^2},
\end{eqnarray}
where $g_V = 1$, $g_A \equiv g_A^{free}=1.2723$,  $(\mu_p - \mu_n) =
4.7$, and the cutoff parameters $\Lambda_V = 850$ MeV and $\Lambda_A =
1086$ MeV.

Then, the total nuclear matrix element \nme~is written as
\begin{equation}
M^{0\nu} =  M_{\rm GT}^{0\nu} - \frac{g_V^2}{g_A^2}  M_{\rm F}^{0\nu}
+  M_{\rm T}^{0\nu}~~.
\label{nme00nu}
\end{equation}

The expression in Eq.~(\ref{M0nu}) cannot be easily calculated within
the nuclear shell model because of the computational complexity of
calculating a large number of intermediate states (the Lanczos
strength-function method \cite{Caurier05} can be applied only for the
single-$\beta$-decay process).
Therefore, most SM calculations resort to the closure approximation,
which is based on the observation that the relative momentum $q$ of
the neutrino, appearing in the propagator of Eq.~(\ref{neutpot}), is of
the order of 100-200 MeV~\cite{Engel17}, and the excitation energies
of the nuclei involved in the transition are of the order of 10
MeV~\cite{Senkov13}.
On these grounds, the energies of the intermediate states appearing in
Eq.~(\ref{neutpot}), may be replaced by an average value
$E_k-(E_i+E_f)/2 \rightarrow \langle E \rangle$, that leads to a
simpler form of both Eqs.~(\ref{M0nu}) and~(\ref{neutpot}).
Consequently, $M_\alpha^{0\nu}$ can be re-written in terms of the
two-body transition-density matrix elements $\langle f |
a^{\dagger}_{p}a_{n} a^{\dagger}_{p^\prime} a_{n^\prime} | i \rangle$
as
\begin{eqnarray}
M_\alpha^{0\nu}& =  & \sum_{j_n j_{n^\prime} j_p j_{p^\prime}}
\langle f | a^{\dagger}_{p}a_{n} a^{\dagger}_{p^\prime} a_{n^\prime} 
| i \rangle \nonumber \\
~ & ~& \times  \left< j_p  j_{p^\prime} \mid \tau^-_{1}
\tau^-_{2} \Theta_\alpha \mid  j_n j_{n^\prime}
       \right>~, \label{M0nuapp}
\end{eqnarray}
and the neutrino potentials become
\begin{equation}
H_{\alpha}(r)=\frac {2R}{\pi} \int_{0}^{\infty} \frac {j_{n_{\alpha}}(qr)
  h_{\alpha}(q^2)qdq}{q+\left< E \right>}~.
\label{neutpotapp}
\end{equation}
As in most SM calculations, we adopt the closure approximation to
define the $\Theta$ operators given in
Eqs.~(\ref{operatorGT})--(\ref{operatorT}), and take the average
energy $\left< E \right>=11.2$ MeV from the evaluation of
Ref.~\cite{Tomoda91}.
As regards the soundness of the closure approximation to evaluate
\nme, we should point out that in Ref.~\cite{Senkov13} the authors
have performed SM calculations of $^{48}$Ca \zbb~decay both within and
beyond the closure approximation, and found that in the second case the
results are $\sim 10\%$ larger.

As mentioned in section \ref{effham}, one needs to consider
short-range correlations when computing the radial matrix elements of
the neutrino potentials $\left < \psi_{nl}(r) |H_{\alpha}|
  \psi_{n^\prime l^\prime}(r) \right>$.

SRC account for the physics that is missing in all models
that expand nuclear wave functions in terms of a truncated
non-correlated SP basis \cite{Bethe71,Kortelainen07}.
This is related to the highly repulsive nature of the short-range
two-nucleon interaction, and in order to carry out our SM calculation,
that is based on effective operators derived from a realistic
potential, we perform a consistent regularization both of the
two-nucleon potential, $V^{NN}$, and the \zbb-decay operator
\cite{Coraggio20d}.

As a matter of fact, the \vlwk~procedure~\cite{Bogner02} renormalizes
the repulsive high-momentum components of the $V^{NN}$ potential
through a unitary transformation $\Omega$.
The latter is an operator which decouples the full momentum space of
the two-nucleon Hamiltonian, $H^{NN}$, into two subspaces; the first
one is associated with the relative-momentum configurations below a
cutoff $\Lambda$ and is specified by a projector operator $P$, the
second one is defined in terms of its complement $Q=\mathbf{1}-P$
\cite{Coraggio20d}.
As unitary transformation, $\Omega$ preserves the physics of the
original potential for the two-nucleon system, namely, the
calculated values of all $NN$ observables are the same as those
reproduced by solving the Schr\"odinger equation for two nucleons
interacting via  $V^{NN}$.

In order to benefit of this procedure, we calculate the two-body \zbb~
operator, $\Theta$, in the momentum space.
Then, $\Theta$ is renormalized using $\Omega$, to provide consistency
with the $V^{NN}$ potential, whose high-momentum (short range)
components are dumped by the introduction of the cutoff $\Lambda$.
The new decay operator is defined as $\Theta_\lwk \equiv P \Omega
\Theta \Omega^{-1} P$ for relative momenta $k < \Lambda$, and is set
to zero for $k > \Lambda$, and its matrix elements are employed as
vertices in the perturbative expansion of the $\hat{\Theta}$ box.

The magnitude of the overall effect of this renormalization procedure
is comparable to using the SRC modeled by the Unitary Correlation
Operator Method~\cite{Menendez09b}, that is a lighter softening of
\nme~with respect to the one provided by Jastrow type
SRC~\cite{Coraggio20d}.

\section{Results}\label{results}
In this section we present the results of our SM calculations.
First, we compare theoretical and experimental low-energies
spectroscopic properties of the parent and granddaughter nuclei
$^{100}$Mo and $^{100}$Ru, respectively.
We show also the results of the GT$^-$ strength distribution and the
calculated NMEs of the \dbb~decay for $^{100}$Mo and compare them with
the available data.

Then, we calculate the nuclear matrix element of the \zbb~decay and
study the convergence behavior of the effective SM operator we have
derived consistently with \heff.
We also discuss the effects of three-valence-nucleon diagrams which
correct the  Pauli-principle violation introduced in systems with more
than two valence nucleons \cite{Towner87}.

As already mentioned, all the calculations are performed employing
theoretical SP energies and TBMEs obtained from the effective
Hamiltonian $H_{\rm eff}^{[44]}$, whose model space is defined  by
$0f_{5/2},1p_{3/2},1p_{1/2},0g_{9/2}$ proton and
$0g_{7/2},1d_{5/2},1d_{3/2},2s_{1/2}$ neutron orbitals, that can be
found in the Supplemental Material \cite{supplemental2021}.

\subsection{Spectroscopy of $^{100}$Mo and $^{100}$Ru}

\begin{center}
\begin{figure}[H]
\includegraphics[scale=0.22,angle=0]{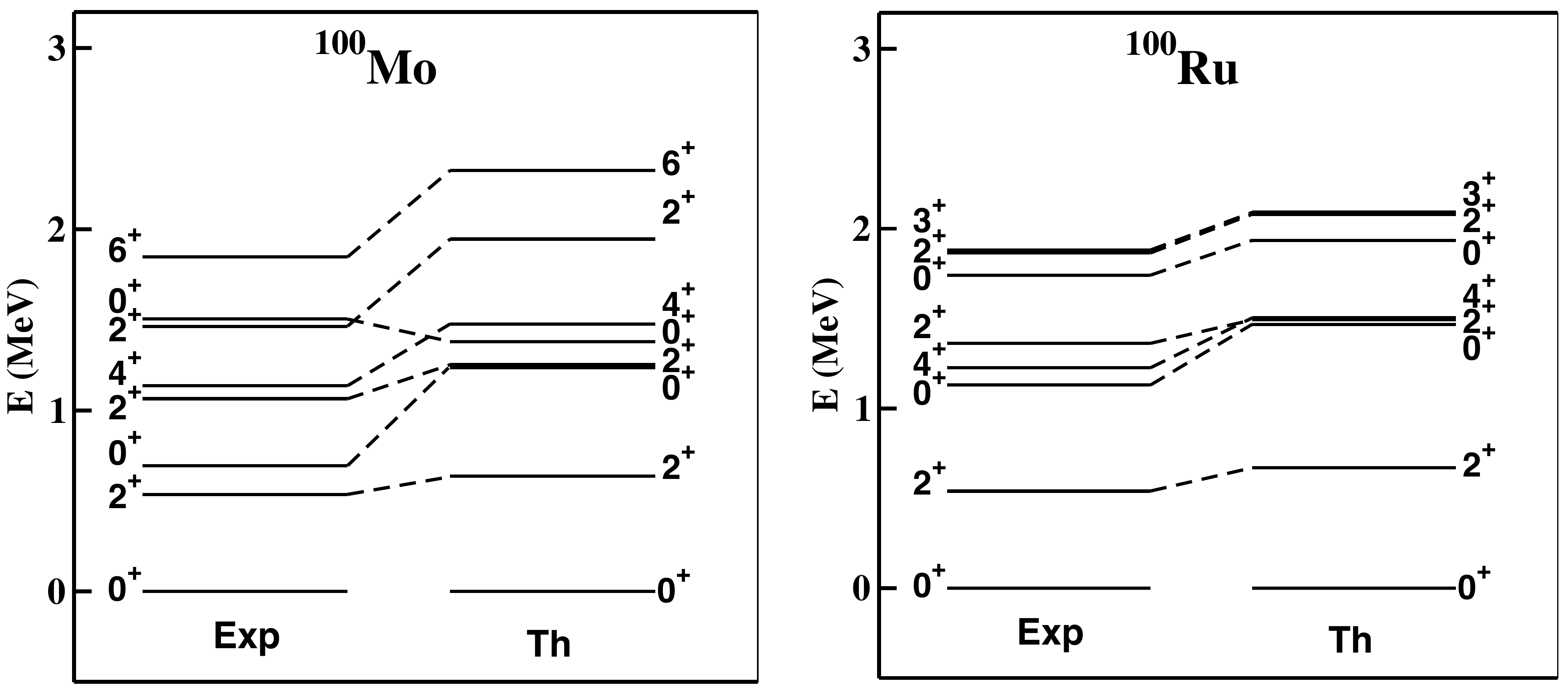}
\caption{Experimental and calculated spectra of $^{100}$Mo and
  $^{100}$Ru up to 2 MeV excitation energy.}
\label{100Mo100Ru}
\end{figure}
\end{center}

In Fig. \ref{100Mo100Ru}, we compare the calculated low-energy spectra
of $^{100}$Mo and $^{100}$Ru, as well as their experimental counterparts.
As can be seen, our \heff~provides a reasonable reproduction of
$^{100}$Mo low-lying states, despite the large number of valence
nucleons involved in the diagonalization of the SM Hamiltonian.
The larger discrepancy between observed and theoretical spectra occurs
for the yrare $J^{\pi}=0^+$ state, which exhibits experimentally a
pronounced collective behavior.
This is also testified by the $B(E2)$ strength between the
$J^{\pi}=0^+_2$ and $J^{\pi}=2^+_1$ levels, that is reported in Table
\ref{E2_100Mo}.
In fact, from the inspection of Table \ref{E2_100Mo}, we see that
there is a general agreement between theoretical and experimental
values, but our calculation fails to reproduce the large $B(E2; 0^+_2
\rightarrow 2^+_1)$.

Once more, it is worth stressing that to calculate the $B(E2)$ strengths
the effective proton/neutron charges have been derived from theory
(see Sec. \ref{effopsec}), without any empirical adjustment, and whose
values can be found in Table \ref{effch}.

\begin{table}[H]
\caption{Proton and neutron effective charges of the electric
  quadrupole operator $E2$.}
\begin{ruledtabular}
\begin{tabular}{cccc}
\label{effch}
$n_a l_a j_a ~ n_b l_b j_b $ &  $\langle a || e_p || b \rangle $ & $n_a l_a j_a ~ n_b l_b j_b $ &  
$\langle a || e_n || b \rangle $ \\
\colrule
 $0f_{5/2}~0f_{5/2}$     & 1.62 & $0g_{7/2}~0g_{7/2}$     & 1.00  \\ 
 $0f_{5/2}~1p_{3/2}$     & 1.45 & $0g_{7/2}~1d_{5/2}$     & 0.73  \\ 
 $0f_{5/2}~1p_{1/2}$     & 1.47 & $0g_{7/2}~1d_{3/2}$     & 0.70  \\ 
 $1p_{3/2}~0f_{5/2}$     & 1.28 & $1d_{5/2}~0g_{7/2}$     & 0.68  \\ 
 $1p_{3/2}~1p_{3/2}$     & 1.20 & $1d_{5/2}~1d_{5/2}$     & 0.47  \\ 
 $1p_{3/2}~1p_{1/2}$     & 1.21 & $1d_{5/2}~1d_{3/2}$     & 0.48  \\ 
 $1p_{1/2}~0f_{5/2}$     & 1.31 & $1d_{5/2}~2s_{1/2}$     & 0.43  \\ 
 $0g_{1/2}~1p_{3/2}$     & 1.22 & $1d_{3/2}~0g_{7/2}$     & 0.66  \\ 
 $0g_{9/2}~0g_{9/2}$     & 1.70 & $1d_{3/2}~1d_{5/2}$     & 0.48  \\ 
             ~                    &    ~   & $1d_{3/2}~1d_{3/2}$     & 0.55  \\ 
             ~                    &    ~   &  $1d_{3/2}~2s_{1/2}$     & 0.50  \\ 
             ~                    &    ~   & $2s_{1/2}~1d_{5/2}$     & 0.43  \\ 
             ~                    &    ~   & $2s_{1/2}~1d_{3/2}$     & 0.50  \\ 
             ~                    &    ~   & $0h_{11/2}~0h_{11/2}$  & 0.79 \\ 
\end{tabular}
\end{ruledtabular}
\end{table}

As regards the low-energy spectrum of $^{100}$Ru, our calculation
provides a satisfactory reproduction of the experiment, and this is
also testified by the comparison of the theoretical $B(E2)$ strengths
with the available data, as reported in Table \ref{E2_100Ru}.

\begin{table}[H]
\caption{Experimental and calculated $B(E2)$ strengths (in $e^2{\rm
  fm}^4$) for $^{100}$Mo, data are taken from Ref. \cite{ensdf}. We
report those for the observed states in Fig. \ref{100Mo100Ru}.}
\begin{ruledtabular}
\begin{tabular}{ccc}
\label{E2_100Mo}
 $J_i \rightarrow J_f $ & $B(E2) _{\rm Expt}$ &  $B(E2) _{\rm Calc}$ \\
\colrule
$2^+_1 \rightarrow 0^+_1$ & $1000 \pm 100$ & 820 \\
$0^+_2 \rightarrow 2^+_1$ & $2500 \pm 100$ & 55 \\
$2^+_2 \rightarrow 0^+_1$ & $17 \pm 1$ & 30 \\
$2^+_2 \rightarrow 2^+_1$ & $1400 \pm 140$ & 800 \\
$2^+_2 \rightarrow 0^+_2$ & $150 \pm 20$ & 540 \\
$4^+_1 \rightarrow 2^+_1$ & $1900 \pm 100$ & 1200 \\
$2^+_3 \rightarrow 2^+_1$ & $8 \pm 2$ &  15 \\
$2^+_3 \rightarrow 0^+_2$ & $400 \pm 100$ & 340 \\
$6^+_1 \rightarrow 4^+_1$ & $2500 \pm 400$ & 1240 \\
\end{tabular}
\end{ruledtabular}
\end{table}
\begin{table}[H]
\caption{Experimental and calculated $B(E2)$ strengths (in $e^2{\rm
    fm}^4$) for $^{100}$Ru, data are taken from Ref. \cite{ensdf}. We
  report those for the observed states in Fig. \ref{100Mo100Ru}.}
\begin{ruledtabular}
\begin{tabular}{ccc}
\label{E2_100Ru}
 $J_i \rightarrow J_f $ & $B(E2) _{\rm Expt}$ &  $B(E2) _{\rm Calc}$ \\
\colrule
$2^+_1 \rightarrow 0^+_1$ & $980 \pm 10$ & 640 \\
$0^+_2 \rightarrow 2^+_1$ & $1000 \pm 140$ & 300 \\
$4^+_1 \rightarrow 2^+_1$ & $1400 \pm 100$ & 980 \\
$2^+_2 \rightarrow 2^+_1$ & $850 \pm 170$ &  570 \\
$2^+_2 \rightarrow 0^+_1$ & $55 \pm 10$ & 50 \\
$2^+_3 \rightarrow 4^+_1$ & $500 \pm 140$ & 90 \\
$2^+_3 \rightarrow 0^+_2$ & $1000 \pm 250$ & 360 \\
\end{tabular}
\end{ruledtabular}
\end{table}

We now proceed to examine the results of the calculation of quantities
that are directly related to the double-$\beta$ decay of $^{100}$Mo.
It is worth pointing out that, because of the proton and neutron model
spaces, the effective GT$^+$ operator consists of one matrix element
that corresponds to the $\pi 0g_{9/2} \rightarrow \nu 0g_{7/2}$ decay,
whose calculated quenching factor is $q=0.454$.
Similarly, the only matrix element of the effective GT$^-$ operator
$\nu 0g_{9/2} \rightarrow \pi 0g_{7/2}$ provides a quenching factor
$q=0.503$.

The reason of a non-hermitian effective GT-decay operator is
threefold; the proton and neutron model spaces we have chosen are
different, the proton-neutron symmetry is broken because the Coulomb
interaction is included in the perturbative expansion, the procedure
that has been followed to derive the effective operators is
non-hermitian \cite{Suzuki95}.

In Table \ref{ME_100Mo} we report the observed and calculated values of
the \nmeds~for the \dbb~decay of $^{100}$Mo from the $J^{\pi}=0^+_1$
ground state (g.s.) to the $^{100}$Ru $J^{\pi}=0^+_1,0^+_2$ states.
For both decays the value of \nmed~ obtained with the bare operator
overestimates the experimental one by a factor $3 \div 4$, but employing the
matrix elements of the effective GT$^+$ and GT$^-$ operators we reach
a result that is in a good agreement with the observed \nmeds.

\begin{table}[H]
  \caption{Experimental \cite{Barabash20}  and calculated \nmeds~(in
    MeV$^{-1}$) for $^{100}$Mo \dbb~decay. The theoretical values are
    obtained employing both the bare (I) and effective (II)
    \dbb~operators.}
\begin{ruledtabular}
\begin{tabular}{cccc}
\label{ME_100Mo}
$^{100}$Mo$\rightarrow^{100}$Ru decay branches & Experiment & I & II \\
\colrule
~ & ~ & ~ & ~ \\
$J^{\pi}=0^+_1\rightarrow J^{\pi}=0^+_1$ & $0.224 \pm 0.002$ & 0.896 & 0.205 \\
$J^{\pi}=0^+_1\rightarrow J^{\pi}=0^+_2$ & $0.182 \pm 0.006$ & 0.479 & 0.109 \\
\colrule
\end{tabular}
\end{ruledtabular}
\end{table}

\begin{center}
\begin{figure}[H]
\includegraphics[scale=0.36,angle=0]{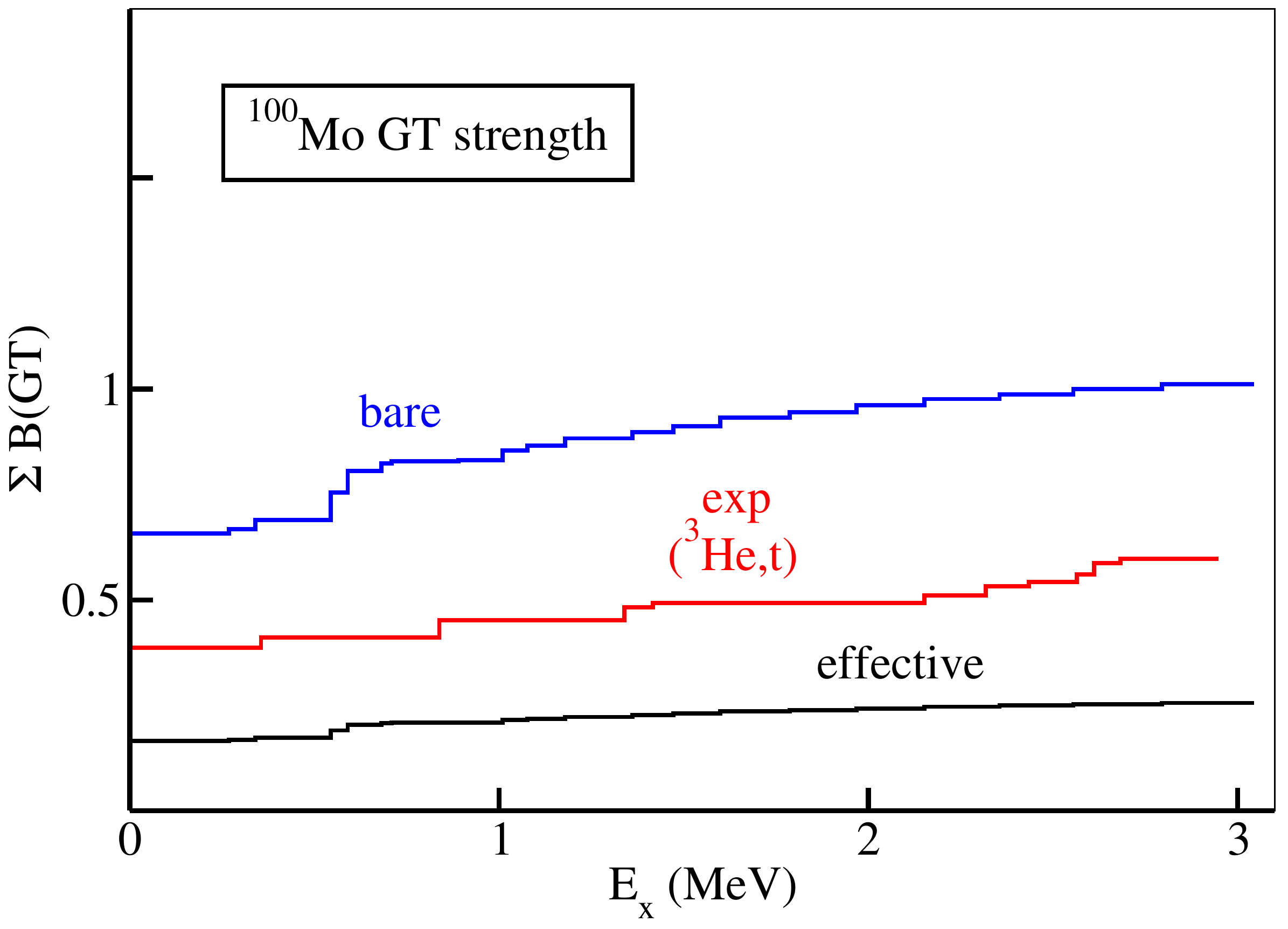}
\caption{Running sums of the $^{100}$Mo $\sum {\rm B(GT)}$ strengths
  as a function of the excitation energy $E_x$ up to 3 MeV.}
\label{100MoGT-}
\end{figure}
\end{center}

In Fig. \ref{100MoGT-}, the calculated $\sum {\rm B(GT)}$ for
$^{100}$Mo are shown as a function of the $^{100}$Tc excitation
energy, and compared with the data reported with a red line
\cite{Thies12c}.
The results obtained with the bare operator are drawn with a blue
line, while those obtained employing the effective GT operator are
plotted a black line.

It can be seen that the distribution obtained using the bare operator
overestimates the observed one, but the quenching induced by the
effective operator provides an underestimation of the values extracted
from the experiment.

Here, it should be reminded that the "experimental" GT strengths
obtained from charge-exchange reactions are not directly observed
data.
The GT strength can be extracted from the GT component of the cross
section at zero degree, following the standard approach in the
distorted-wave Born approximation (DWBA). 

\[
\frac{d\sigma^{GT}(0^\circ)}{d\Omega} = \left (\frac{\mu}{\pi \hbar^2} \right
)^2 \frac{k_f}{k_i} N^{\sigma \tau}_{D}| J_{\sigma \tau} |^2 B({\rm GT})~~,
\]

where $N^{\sigma \tau}_{D}$ is the distortion factor, $| J_{\sigma
  \tau} |$ is the volume integral of the effective $NN$ interaction,
$k_i$ and $k_f$ are the initial and final momenta, respectively, and
$\mu$ is the reduced mass (see formula and description in
Refs. \cite{Puppe12,Frekers13}).
Then, the values of experimental GT strengths are somehow
model-dependent.

\subsection{Neutrinoless double-$\beta$ decay of $^{100}$Mo}

As introduced in Section \ref{outline}, our calculation of
\nme~accounts for the light-neutrino exchange mechanism, the total
nuclear matrix element being expressed as in Eq. (\ref{nme00nu}) and calculated
accordingly to
Eqs. (\ref{operatorGT},\ref{operatorF},\ref{operatorT},\ref{M0nuapp},\ref{neutpotapp}),
namely within the closure approximation.

The perturbative expansion of the \zbb~effective operator \thetaeff~
has been carried out including in the \tbox~ diagrams up to the third
order (see Section \ref{outline}), and a number of intermediate states
which corresponds to oscillator quanta up to $N_{\rm  max}=14$, since
the results are substantially convergent from $N_{\rm max}=12$ on
(see Ref. \cite{Coraggio20a}).

As regards the expansion of \thetaeff~as a function of the $\chi_n$
operators, we stop at $n=2$ since $\chi_3$ depends on the first,
second, and third derivatives of $\hat{\Theta}_0$ and
$\hat{\Theta}_{00}$, as well as on the first and second derivatives of
the $\hat{Q}$ box (see Eq. (\ref{chin})), so $\chi_3$ contribution may
be estimated at least one order of magnitude smaller than the $\chi_2$
one.
Moreover, in Ref \cite{Coraggio20a} we have shown that the
contributions from $\chi_1$ are relevant, while those from $\chi_2$
are almost negligible.

\begin{figure}[h]
\begin{center}
\includegraphics[scale=0.32,angle=0]{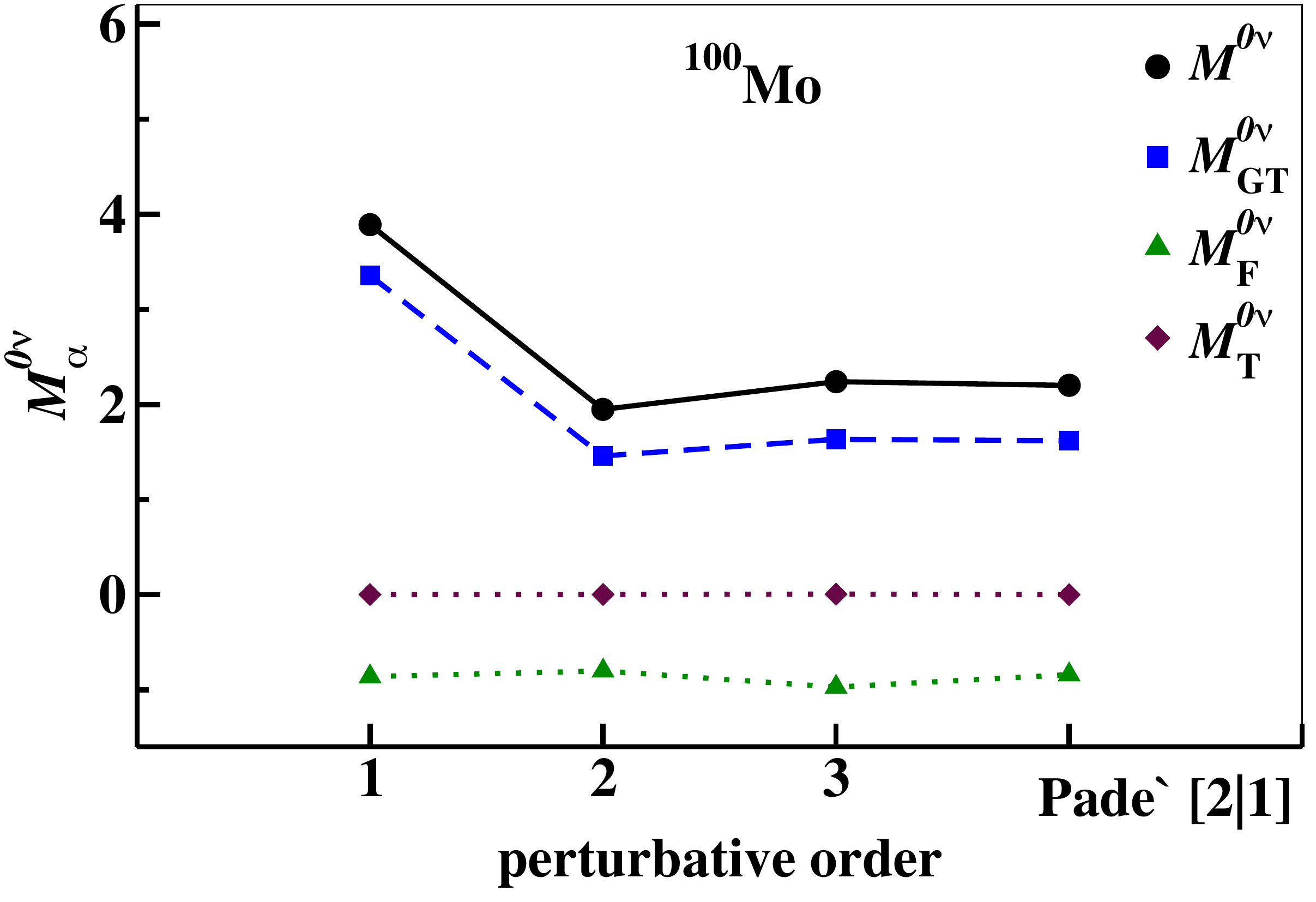}
\caption{\nme~for the decay of the $^{100}$Mo $J^{\pi}=0^+_1$ state to
  the $^{100}$Ru $J^{\pi}=0^+_1$ one, as a function of the
  perturbative order. The green triangles correspond to $M^{0\nu}_{\rm
    F}$, the blue squares to $M^{0\nu}_{\rm GT}$, the magenta diamonds
  to $M^{0\nu}_{\rm T}$, and the black dots to the full \nme.}
\label{100Mo_obo}
\end{center}
\end{figure}

\begin{figure}[h]
\begin{center}
\includegraphics[scale=0.32,angle=0]{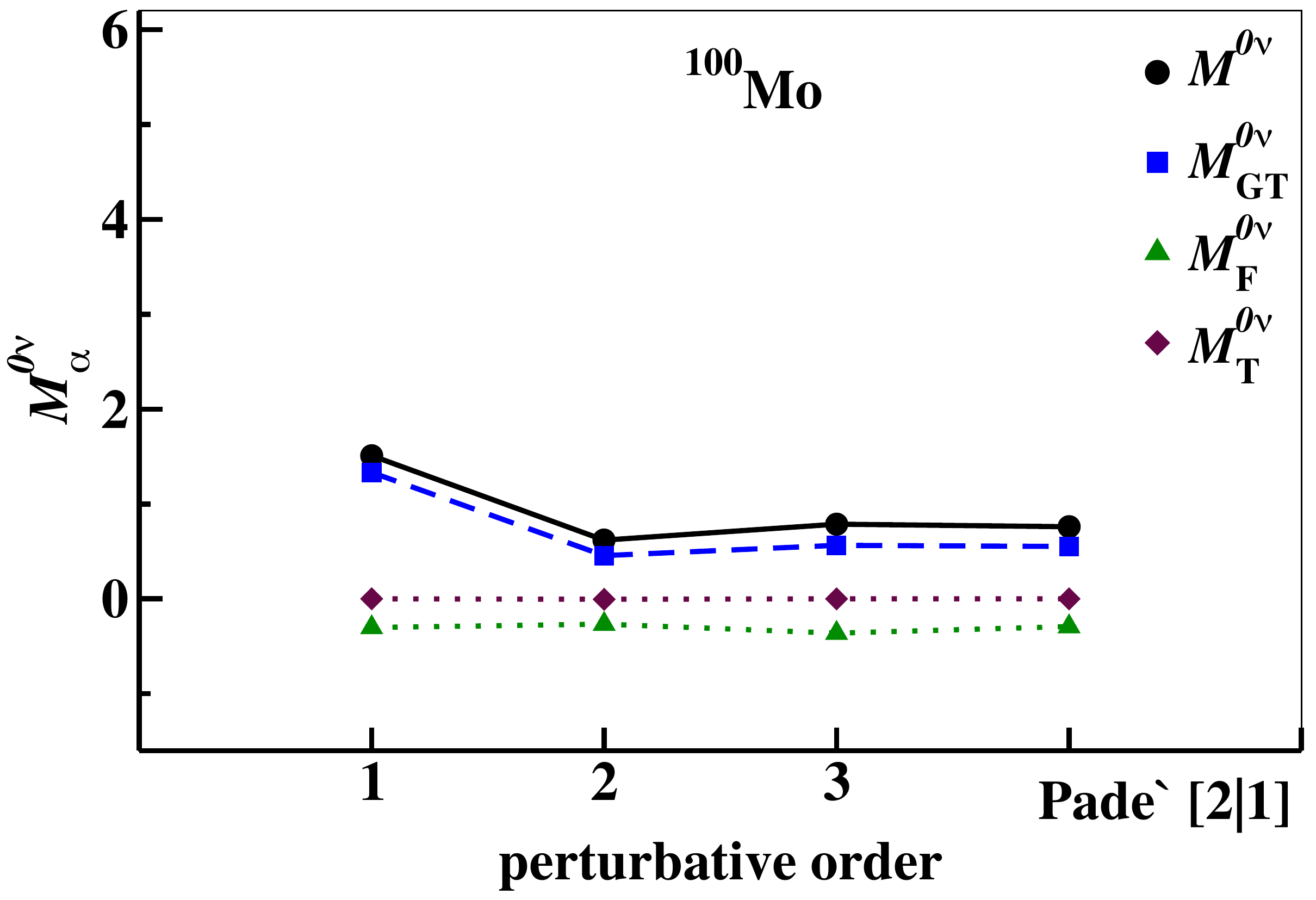}
\caption{Same as in Fig. \ref{100Mo_obo}, but for the decay of the
  $^{100}$Mo $J^{\pi}=0^+_1$ state to the $^{100}$Ru $J^{\pi}=0^+_2$
  one.}
\label{100Mo_obo-II}
\end{center}
\end{figure}

First, we focus on the results of the order-by-order convergence
behavior by reporting in Figs. \ref{100Mo_obo},\ref{100Mo_obo-II} the
calculated values of \nme, $M^{0\nu}_{\rm GT}$, $M^{0\nu}_{\rm F}$,
and $M^{0\nu}_{\rm T}$ for both the decay of the $^{100}$Mo
$J^{\pi}=0^+_1$ state to the $^{100}$Ru $J^{\pi}=0^+_1,0^+_2$ ones,
respectively,  from first- up to third-order in perturbation theory.
As an indicator of the quality of the perturbative behavior
\cite{Baker70}, we also report the value of their Pad\'e approximant
$[2|1]$.
We also point out that the same scale has been adopted in both
figures.

As in other decays we have studied in our previous work
\cite{Coraggio20a}, the perturbative behavior is driven by the
Gamow-Teller component, since the Fermi matrix element $M^{0\nu}_{\rm
  F}$ is weakly affected by the renormalization procedure, and
$M^{0\nu}_{\rm T}$ is almost negligible.
We observe a perturbative pattern of the calculated \nme~of $^{100}$Mo
that is better than the ones we have found for $^{48}$Ge, $^{76}$Ge,
$^{82}$Se, $^{130}$Te, and $^{136}$Xe \zbb~decays, which have been
calculated within the same approach \cite{Coraggio20a}.
In fact, here the difference between second- and third-order
results is about $13\%$ and $21\%$ for the decay of the $^{100}$Mo
$J^{\pi}=0^+_1$ state to the $^{100}$Ru $J^{\pi}=0^+_1$ and
$J^{\pi}=0^+_2$ ones, respectively.

\begin{table}[ht]
  \caption{Calculated values of \nme~for the decay of $^{100}$Mo
    g.s. state to the yrast and yrare $J^{\pi}=0^+$ states of
    $^{100}$Ru.}
\begin{ruledtabular}
\begin{tabular}{lllll}
\label{NMEtab}
 ~ $0^+_1 \rightarrow  0^+_1$ & ~ & ~ & ~ & ~ \\
 ~ & $M^{0\nu}_{\rm GT}$ & $M^{0\nu}_{\rm F}$ & $M^{0\nu}_{\rm T}$  & \nme \\
\colrule
 Present work (I)                    & 3.418 & -0.878 & 0.002 & 3.962 \\
 Present work (II)                   & 1.634 & -0.970 & 0.007 & 2.240 \\
  IBM-2 \cite{Barea15}           & 3.73   &  -0.48  &   0.19  & 4.22 \\
  EDF \cite{Vaquero13}          & 5.361 & -1.986 &     ~     & 6.588 \\
  BMF-CDFT \cite{Yao15}      & ~ & ~ &             ~             & 10.91 \\
  pnQRPA \cite{Simkovic13}  & 4.950 & -2.367 & -0.571 & 5.850 \\
  pnQRPA \cite{Hyvarinen15} &  3.13 &  -1.03  &  -0.26  & 3.90 \\
\colrule
 ~ & ~ & ~ & ~ & ~ \\
 ~ $0^+_1 \rightarrow  0^+_2$ & ~ & ~ & ~ & ~ \\
 ~ & $M^{0\nu}_{\rm GT}$ & $M^{0\nu}_{\rm F}$ & $M^{0\nu}_{\rm T}$  & \nme \\
\colrule
 Present work (I)        & 1.344 & -0.308 & 0.001 & 1.535 \\
 Present work (II)        & 0.564 & -0.361 & 0.001 & 0.788 \\
IBM-2 \cite{Barea15} &  0.99   &  -0.13  &  0.05  & 1.12 
\end{tabular}
\end{ruledtabular}
\end{table}

In Table \ref{NMEtab} the values of \nme, which we have calculated by
using both the bare operator -- namely without condidering neither SRC
nor renormalizations due to the truncation of the model space -- and
\thetaeff, have been reported, as well as their Gamow-Teller, Fermi, and
tensor components.
Our results are also compared with those obtained employing other
nuclear models, such as the interacting boson model with isospin
restoration (IBM-2) \cite{Barea15}, the energy density functional
method including deformation and pairing fluctuations (EDF)
\cite{Vaquero13}, the beyond-mean-field covariant density functional
theory (BMF-CDFT) \cite{Yao15}, and quasiparticle random-phase
approximation with isospin symmetry restoration (pnQRPA)
\cite{Simkovic13,Hyvarinen15}.

The SM results obtained with the bare \zbb~ operator (I) can be better
compared with other nuclear models, since in the latter no effective
operator has been considered, and we see that our \nmes~are close to
those in Refs. \cite{Barea15,Hyvarinen15}, where the IBM-2 and pnQRPA
models have been employed, respectively.
The other calculations provide \nmes~ that are much larger than our
result, and it is worth pointing out that different choices of the
parameters for pnQRPA calculations may lead to a remarkable difference
of the calculated \nmes~\cite{Simkovic13,Hyvarinen15}.

The action of the effective operator \thetaeff~ quenches the value of
the two \nmes~ by a factor about one half, whose effect is smaller
than accounting for the quenching factor of the axial coupling
constant $g_A$ that comes out from the calculated effective GT$^{\pm}$
operator, which is about $q=0.5$.

These considerations are related to the question if one should relate
the derivation of the effective one-body GT operator
\cite{Coraggio19a} with the renormalization of the two-body GT
component of the \zbb~operator.
As a matter of fact, this issue has a considerable impact on the
detectability of \zbb~process \cite{Suhonen17a,Suhonen17b}.

\begin{figure}[H]
\begin{center}
\includegraphics[scale=0.32,angle=0]{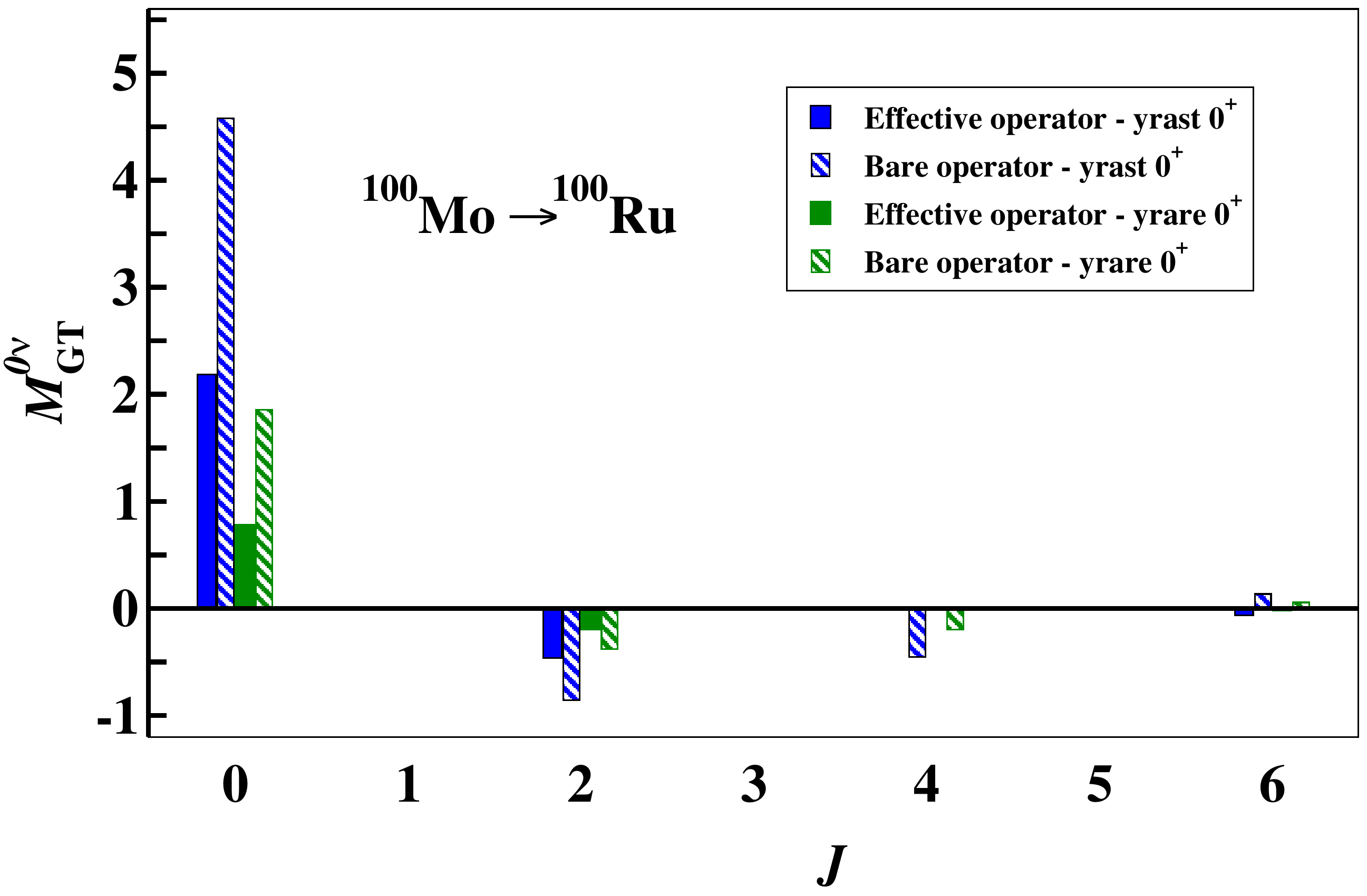}
\caption{Contributions from pairs of decaying neutrons with given
  $J^{\pi}$ to $M^{0\nu}_{\rm GT}$ for $^{100}$Mo \zbb~decay. The bars
  filled in blue corresponds to the results obtained with $\Theta_{\rm
    eff}$, those in dashed blue to the ones calculated with bare
  operator}
\label{100MoGT-jj}
\end{center}
\end{figure}

\begin{figure}[h]
\begin{center}
\includegraphics[scale=0.32,angle=0]{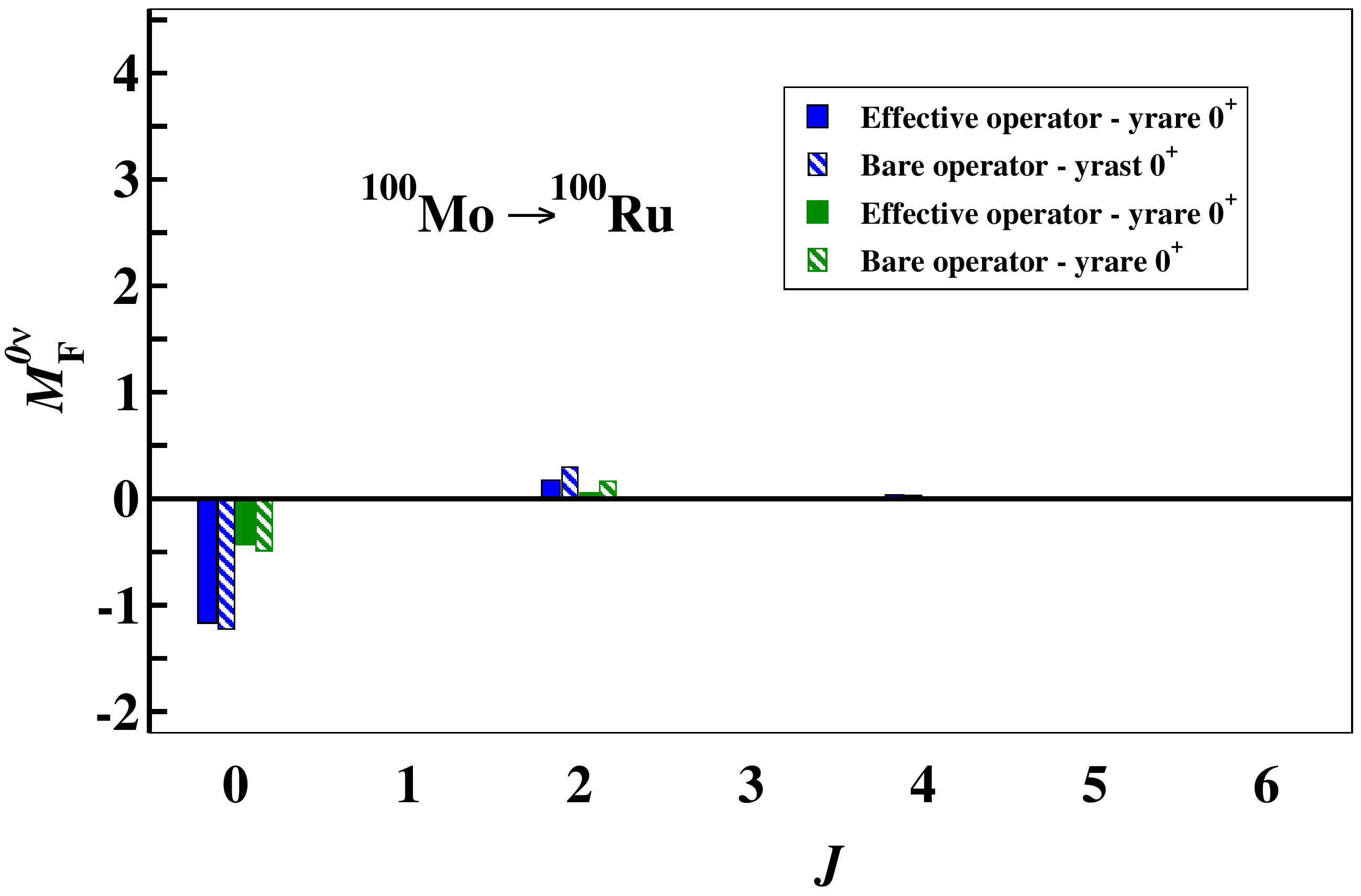}
\caption{Same as in Fig. \ref{100MoGT-jj}, but for $M^{0\nu}_{\rm F}$.}
\label{100MoF-jj}
\end{center}
\end{figure}

To complete our discussion about the \nmes, we show in
Figs. \ref{100MoGT-jj}, \ref{100MoF-jj} the results of the
decomposition of $M^{0\nu}_{\rm GT}$ and $M^{0\nu}_{\rm F}$,
respectively, in terms of the contributions from the decaying pair of
neutrons coupled to a given angular momentum and parity $J^{\pi}$,
both for the decay to the $^{100}$Ru ground (blue columns) and yrare
$J^{\pi}=0^+$ (green columns) states.

We report the contributions obtained by employing both the effective
\zbb-decay~ operator $\Theta_{\rm eff}$ (colour filled columns) and the
bare one (dashed filled columns).

The results of the decomposition of $M^{0\nu}_{\rm F}$ confirm the
irrelevance of the renormalization procedure, and exhibit the
dominance of the $J^{\pi}=0^+$ component.

As regards $M^{0\nu}_{\rm GT}$, as it should be expected, each
$J^{\pi}$ contribution  calculated employing \thetaeff~ is much
smaller than the one obtained with the bare \zbb-decay operator.
The main contributions, both employing effective and bare operators,
correspond to the $J^{\pi}=0^+,2^+$ components, being opposite in
sign, and a non-negligible role is played by the $J^{\pi}=4^+$
component too.

\section{Summary and outlook}\label{conclusions}
This work is the first attempt to calculate double-$\beta$ decay of
$^{100}$Mo into $^{100}$Ru by way of the nuclear shell model.

Our study has consisted first in verifying the ability of the tools we
have chosen, namely the model space and the shell-model effective
Hamiltonian and decay operators, to reproduce the experimental
spectroscopic properties of $^{100}$Mo,Ru -- excitation spectra
and $B(E2)$ strengths that are related to the collective behavior of
these systems -- as well as the nuclear matrix elements \nmed~of the
\dbb-decay and the GT strengths obtained from charge-exchange
reactions.
Then, after having tested and shown the degree of reliability of our
wave functions, we have calculated the nuclear matrix elements \nmes~
of the \zbb-decay of the $^{100}$Mo ground state to the yrast and
yrare $J^{\pi}=0^+$ states of $^{100}$Ru.

An important feature of our work is that shell-model effective
Hamiltonians and decay operators have been derived by way of many-body
perturbation theory, starting from a high-precision realistic
potential CD-Bonn \cite{Machleidt01b}.
Such an approach has been previously applied to study the 
$^{48}$Ca$\rightarrow^{48}$Ti, $^{76}$Ge$\rightarrow^{76}$Se,
$^{82}$Se$\rightarrow^{82}$Kr, $^{130}$Te$\rightarrow^{130}$Xe, and
$^{136}$Xe$\rightarrow^{136}$Ba decays
\cite{Coraggio17a,Coraggio19a,Coraggio20a}.

The comparison of our results with the available data seems to
indicate that the realistic shell model can quantitatively describe
most of the spectroscopy (low-lying excitation spectra,
electromagnetic transition strengths) of $^{100}$Mo,Ru and also their
$\beta$-decay properties (nuclear matrix elements of \dbb~decay, GT
strengths from charge-exchange reactions) without resorting to
empirical adjustments of \heff, effective charges, or quenching the
axial coupling constant.
This should provide support to our approach for the prediction of the
\nmes~for the \zbb~decay of $^{100}$Mo, within the
light-neutrino-exchange channel, that is a conjugation of the action
of shell-model wave functions, emerging from the diagonalization of
\heff, and effective decay operators, which are constructed
consistently with \heff.

We have also compared our results for the \zbb~decay of $^{100}$Mo
with those obtained employing other nuclear methods, leading to some
relevant observations.
To this end, we have considered the results we obtain employing both
the bare \zbb~ operator -- namely without any sort of normalization --
and the effective operator derived theoretically.
The \nmes~ we calculate with the bare operator are important for a fair
comparison with other nuclear models, since the latter do not employ
any effective operator which accounts for the truncation of the
Hilbert space.

First, it can be noticed that our results with the bare operator are
consistent with recent calculations performed with IBM-2 \cite{Barea15} and
pnQRPA \cite{Hyvarinen15}, whereas the results obtained within EDF
\cite{Vaquero13} and BMF-CDFT \cite{Yao15} approaches, as well as
pnQRPA calculations performed by {\v{S}}imkovic {\it et al.}
\cite{Simkovic13}, provide larger values of \nmes.

Second, as in our previous study \cite{Coraggio20a}, the effect of the
renormalization of the \zbb-decay operator, with respect to the
truncation of the full Hilbert space to the shell-model one, is
smaller than the one obtained for the \dbb-decay one.

These results may be a valuable asset for the community that is
involved with the experimental detection of the $^{100}$Mo \zbb-decay,
since this is the first time a microscopic calculation has been
performed of the \nmes~ of the $^{100}$Mo ground-state decay to the
two lowest-in-energy $J^{\pi}=0^+$ states of $^{100}$Ru.

Our future program to upgrade the study of nuclei with mass $A \approx
100$ which are candidates to \zbb-decay is twofold.

On one side, we plan to start from nuclear forces that own a firm link
with QCD, namely we will construct effective shell-model Hamiltonians
and decay-operators from two- and three-body potentials derived within
the framework of chiral perturbation theory
\cite{Fukui18,Ma19,Coraggio20e}. 

This step will allow us:

\begin{enumerate}
\item[a)] to evaluate the dependence of the predictions for \nmes~ on
  the nuclear potential that is employed in a nuclear structure calculation;
\item[b)] to benchmark our results with those obtained with {\it ab
    initio} calculations \cite{Yao20a,Novario21,Belley21};  
\item[c)] to consider the contribution of the two-body meson-exchange
  corrections to the electroweak currents, originated from
  sub-nucleonic degrees of freedom, that can be consistently tackled
  employing nuclear chiral potentials.
\end{enumerate}

On the other side, we are currently exploring the possibility to
employ larger model spaces, that would account better for the
low-energy collective behavior of nuclei with mass $A \approx 100$.
This would provide major informations about the connection between the
calculated values of the \nmes~and the dimension of the model space,
and how theoretical effective decay-operators can compensate
and reduce this dependence.

These goals are computationally challenging, but we are confident that
our current efforts may lead in a close future to a first set of
preliminary results.

\section*{Acknowledgements}
We acknowledge the CINECA award under the ISCRA initiative through the
INFN-CINECA agreement, for the availability of high performance
computing resources and support.
We acknowledge PRACE for awarding access to the Fenix Infrastructure
resources, which are partially funded from the European Union's
Horizon 2020 research and innovation programme through the ICEI
project under the grant agreement No. 800858.
G. De Gregorio acknowledges the support by the funding program
``VALERE'' of Universit\`a degli Studi della Campania ``Luigi
Vanvitelli''.

\bibliographystyle{apsrev}
\bibliography{biblio.bib}

\begin{thebibliography}{99}
\expandafter\ifx\csname natexlab\endcsname\relax\def\natexlab#1{#1}\fi
\expandafter\ifx\csname bibnamefont\endcsname\relax
  \def\bibnamefont#1{#1}\fi
\expandafter\ifx\csname bibfnamefont\endcsname\relax
  \def\bibfnamefont#1{#1}\fi
\expandafter\ifx\csname citenamefont\endcsname\relax
  \def\citenamefont#1{#1}\fi
\expandafter\ifx\csname url\endcsname\relax
  \def\url#1{\texttt{#1}}\fi
\expandafter\ifx\csname urlprefix\endcsname\relax\def\urlprefix{URL }\fi
\providecommand{\bibinfo}[2]{#2}
\providecommand{\eprint}[2][]{\url{#2}}

\bibitem[{\citenamefont{Fukuda et~al.}(1998)\citenamefont{Fukuda, Hayakawa,
  Ichihara, Inoue, Ishihara, Ishino, Itow, Kajita, Kameda, Kasuga
  et~al.}}]{Fukuda98}
\bibinfo{author}{\bibfnamefont{Y.}~\bibnamefont{Fukuda}},
  \bibinfo{author}{\bibfnamefont{T.}~\bibnamefont{Hayakawa}},
  \bibinfo{author}{\bibfnamefont{E.}~\bibnamefont{Ichihara}},
  \bibinfo{author}{\bibfnamefont{K.}~\bibnamefont{Inoue}},
  \bibinfo{author}{\bibfnamefont{K.}~\bibnamefont{Ishihara}},
  \bibinfo{author}{\bibfnamefont{H.}~\bibnamefont{Ishino}},
  \bibinfo{author}{\bibfnamefont{Y.}~\bibnamefont{Itow}},
  \bibinfo{author}{\bibfnamefont{T.}~\bibnamefont{Kajita}},
  \bibinfo{author}{\bibfnamefont{J.}~\bibnamefont{Kameda}},
  \bibinfo{author}{\bibfnamefont{S.}~\bibnamefont{Kasuga}},
  \bibnamefont{et~al.} (\bibinfo{collaboration}{Super-Kamiokande
  Collaboration}), \bibinfo{journal}{Phys. Rev. Lett.}
  \textbf{\bibinfo{volume}{81}}, \bibinfo{pages}{1562} (\bibinfo{year}{1998}).

\bibitem[{\citenamefont{Ahmad et~al.}(2001)\citenamefont{Ahmad, Allen,
  Andersen, Anglin, B\"uhler, Barton, Beier, Bercovitch, Bigu, Biller
  et~al.}}]{Ahmad01}
\bibinfo{author}{\bibfnamefont{Q.~R.} \bibnamefont{Ahmad}},
  \bibinfo{author}{\bibfnamefont{R.~C.} \bibnamefont{Allen}},
  \bibinfo{author}{\bibfnamefont{T.~C.} \bibnamefont{Andersen}},
  \bibinfo{author}{\bibfnamefont{J.~D.} \bibnamefont{Anglin}},
  \bibinfo{author}{\bibfnamefont{G.}~\bibnamefont{B\"uhler}},
  \bibinfo{author}{\bibfnamefont{J.~C.} \bibnamefont{Barton}},
  \bibinfo{author}{\bibfnamefont{E.~W.} \bibnamefont{Beier}},
  \bibinfo{author}{\bibfnamefont{M.}~\bibnamefont{Bercovitch}},
  \bibinfo{author}{\bibfnamefont{J.}~\bibnamefont{Bigu}},
  \bibinfo{author}{\bibfnamefont{S.}~\bibnamefont{Biller}},
  \bibnamefont{et~al.} (\bibinfo{collaboration}{SNO Collaboration}),
  \bibinfo{journal}{Phys. Rev. Lett.} \textbf{\bibinfo{volume}{87}},
  \bibinfo{pages}{071301} (\bibinfo{year}{2001}).

\bibitem[{\citenamefont{Falcone and Tramontano}(2001)}]{Falcone01}
\bibinfo{author}{\bibfnamefont{D.}~\bibnamefont{Falcone}} \bibnamefont{and}
  \bibinfo{author}{\bibfnamefont{F.}~\bibnamefont{Tramontano}},
  \bibinfo{journal}{Phys. Rev. D} \textbf{\bibinfo{volume}{64}},
  \bibinfo{pages}{077302} (\bibinfo{year}{2001}).

\bibitem[{\citenamefont{Mohapatra and Smirnov}(2006)}]{Mohapatra06}
\bibinfo{author}{\bibfnamefont{R.}~\bibnamefont{Mohapatra}} \bibnamefont{and}
  \bibinfo{author}{\bibfnamefont{A.}~\bibnamefont{Smirnov}},
  \bibinfo{journal}{Annu. Rev. Nucl. Part. Sci.} \textbf{\bibinfo{volume}{56}},
  \bibinfo{pages}{569} (\bibinfo{year}{2006}).

\bibitem[{\citenamefont{Dell'Oro et~al.}(2016)\citenamefont{Dell'Oro, Marcocci,
  Viel, and Vissani}}]{DellOro15}
\bibinfo{author}{\bibfnamefont{S.}~\bibnamefont{Dell'Oro}},
  \bibinfo{author}{\bibfnamefont{S.}~\bibnamefont{Marcocci}},
  \bibinfo{author}{\bibfnamefont{M.}~\bibnamefont{Viel}}, \bibnamefont{and}
  \bibinfo{author}{\bibfnamefont{F.}~\bibnamefont{Vissani}},
  \bibinfo{journal}{Adv. High Energy Phys.} \textbf{\bibinfo{volume}{2016}},
  \bibinfo{pages}{2162659} (\bibinfo{year}{2016}).

\bibitem[{\citenamefont{Kotila and Iachello}(2012)}]{Kotila12}
\bibinfo{author}{\bibfnamefont{J.}~\bibnamefont{Kotila}} \bibnamefont{and}
  \bibinfo{author}{\bibfnamefont{F.}~\bibnamefont{Iachello}},
  \bibinfo{journal}{Phys. Rev. C} \textbf{\bibinfo{volume}{85}},
  \bibinfo{pages}{034316} (\bibinfo{year}{2012}).

\bibitem[{\citenamefont{Kotila and Iachello}(2013)}]{Kotila13}
\bibinfo{author}{\bibfnamefont{J.}~\bibnamefont{Kotila}} \bibnamefont{and}
  \bibinfo{author}{\bibfnamefont{F.}~\bibnamefont{Iachello}},
  \bibinfo{journal}{Phys. Rev. C} \textbf{\bibinfo{volume}{87}},
  \bibinfo{pages}{024313} (\bibinfo{year}{2013}).

\bibitem[{\citenamefont{Avignone et~al.}(2008)\citenamefont{Avignone, Elliott,
  and Engel}}]{Avignone08}
\bibinfo{author}{\bibfnamefont{F.~T.} \bibnamefont{Avignone}},
  \bibinfo{author}{\bibfnamefont{S.~R.} \bibnamefont{Elliott}},
  \bibnamefont{and} \bibinfo{author}{\bibfnamefont{J.}~\bibnamefont{Engel}},
  \bibinfo{journal}{Rev. Mod. Phys.} \textbf{\bibinfo{volume}{80}},
  \bibinfo{pages}{481} (\bibinfo{year}{2008}).

\bibitem[{\citenamefont{Tanabashi et~al.}(2018)\citenamefont{Tanabashi,
  Hagiwara, Hikasa, Nakamura, Sumino, Takahashi, Tanaka, Agashe, Aielli, Amsler
  et~al.}}]{PDG18}
\bibinfo{author}{\bibfnamefont{M.}~\bibnamefont{Tanabashi}},
  \bibinfo{author}{\bibfnamefont{K.}~\bibnamefont{Hagiwara}},
  \bibinfo{author}{\bibfnamefont{K.}~\bibnamefont{Hikasa}},
  \bibinfo{author}{\bibfnamefont{K.}~\bibnamefont{Nakamura}},
  \bibinfo{author}{\bibfnamefont{Y.}~\bibnamefont{Sumino}},
  \bibinfo{author}{\bibfnamefont{F.}~\bibnamefont{Takahashi}},
  \bibinfo{author}{\bibfnamefont{J.}~\bibnamefont{Tanaka}},
  \bibinfo{author}{\bibfnamefont{K.}~\bibnamefont{Agashe}},
  \bibinfo{author}{\bibfnamefont{G.}~\bibnamefont{Aielli}},
  \bibinfo{author}{\bibfnamefont{C.}~\bibnamefont{Amsler}},
  \bibnamefont{et~al.} (\bibinfo{collaboration}{Particle Data Group}),
  \bibinfo{journal}{Phys. Rev. D} \textbf{\bibinfo{volume}{98}},
  \bibinfo{pages}{030001} (\bibinfo{year}{2018}).

\bibitem[{\citenamefont{Pastore et~al.}(2018)\citenamefont{Pastore, Carlson,
  Cirigliano, Dekens, Mereghetti, and Wiringa}}]{Pastore18}
\bibinfo{author}{\bibfnamefont{S.}~\bibnamefont{Pastore}},
  \bibinfo{author}{\bibfnamefont{J.}~\bibnamefont{Carlson}},
  \bibinfo{author}{\bibfnamefont{V.}~\bibnamefont{Cirigliano}},
  \bibinfo{author}{\bibfnamefont{W.}~\bibnamefont{Dekens}},
  \bibinfo{author}{\bibfnamefont{E.}~\bibnamefont{Mereghetti}},
  \bibnamefont{and} \bibinfo{author}{\bibfnamefont{R.~B.}
  \bibnamefont{Wiringa}}, \bibinfo{journal}{Phys. Rev. C}
  \textbf{\bibinfo{volume}{97}}, \bibinfo{pages}{014606}
  (\bibinfo{year}{2018}).

\bibitem[{\citenamefont{Cirigliano et~al.}(2018)\citenamefont{Cirigliano,
  Dekens, de~Vries, Graesser, Mereghetti, Pastore, and van
  Kolck}}]{Cirigliano18}
\bibinfo{author}{\bibfnamefont{V.}~\bibnamefont{Cirigliano}},
  \bibinfo{author}{\bibfnamefont{W.}~\bibnamefont{Dekens}},
  \bibinfo{author}{\bibfnamefont{J.}~\bibnamefont{de~Vries}},
  \bibinfo{author}{\bibfnamefont{M.~L.} \bibnamefont{Graesser}},
  \bibinfo{author}{\bibfnamefont{E.}~\bibnamefont{Mereghetti}},
  \bibinfo{author}{\bibfnamefont{S.}~\bibnamefont{Pastore}}, \bibnamefont{and}
  \bibinfo{author}{\bibfnamefont{U.}~\bibnamefont{van Kolck}},
  \bibinfo{journal}{Phys. Rev. Lett.} \textbf{\bibinfo{volume}{120}},
  \bibinfo{pages}{202001} (\bibinfo{year}{2018}).

\bibitem[{\citenamefont{Cirigliano et~al.}(2019)\citenamefont{Cirigliano,
  Dekens, de~Vries, Graesser, Mereghetti, Pastore, Piarulli, van Kolck, and
  Wiringa}}]{Cirigliano19}
\bibinfo{author}{\bibfnamefont{V.}~\bibnamefont{Cirigliano}},
  \bibinfo{author}{\bibfnamefont{W.}~\bibnamefont{Dekens}},
  \bibinfo{author}{\bibfnamefont{J.}~\bibnamefont{de~Vries}},
  \bibinfo{author}{\bibfnamefont{M.~L.} \bibnamefont{Graesser}},
  \bibinfo{author}{\bibfnamefont{E.}~\bibnamefont{Mereghetti}},
  \bibinfo{author}{\bibfnamefont{S.}~\bibnamefont{Pastore}},
  \bibinfo{author}{\bibfnamefont{M.}~\bibnamefont{Piarulli}},
  \bibinfo{author}{\bibfnamefont{U.}~\bibnamefont{van Kolck}},
  \bibnamefont{and} \bibinfo{author}{\bibfnamefont{R.~B.}
  \bibnamefont{Wiringa}}, \bibinfo{journal}{Phys. Rev. C}
  \textbf{\bibinfo{volume}{100}}, \bibinfo{pages}{055504}
  (\bibinfo{year}{2019}).

\bibitem[{\citenamefont{Yao et~al.}(2020)\citenamefont{Yao, Bally, Engel,
  Wirth, Rodr\'{\i}guez, and Hergert}}]{Yao20a}
\bibinfo{author}{\bibfnamefont{J.~M.} \bibnamefont{Yao}},
  \bibinfo{author}{\bibfnamefont{B.}~\bibnamefont{Bally}},
  \bibinfo{author}{\bibfnamefont{J.}~\bibnamefont{Engel}},
  \bibinfo{author}{\bibfnamefont{R.}~\bibnamefont{Wirth}},
  \bibinfo{author}{\bibfnamefont{T.~R.} \bibnamefont{Rodr\'{\i}guez}},
  \bibnamefont{and} \bibinfo{author}{\bibfnamefont{H.}~\bibnamefont{Hergert}},
  \bibinfo{journal}{Phys. Rev. Lett.} \textbf{\bibinfo{volume}{124}},
  \bibinfo{pages}{232501} (\bibinfo{year}{2020}).

\bibitem[{\citenamefont{Novario et~al.}(2021)\citenamefont{Novario, Gysbers,
  Engel, Hagen, Jansen, Morris, Navr\'atil, Papenbrock, and
  Quaglioni}}]{Novario21}
\bibinfo{author}{\bibfnamefont{S.}~\bibnamefont{Novario}},
  \bibinfo{author}{\bibfnamefont{P.}~\bibnamefont{Gysbers}},
  \bibinfo{author}{\bibfnamefont{J.}~\bibnamefont{Engel}},
  \bibinfo{author}{\bibfnamefont{G.}~\bibnamefont{Hagen}},
  \bibinfo{author}{\bibfnamefont{G.~R.} \bibnamefont{Jansen}},
  \bibinfo{author}{\bibfnamefont{T.~D.} \bibnamefont{Morris}},
  \bibinfo{author}{\bibfnamefont{P.}~\bibnamefont{Navr\'atil}},
  \bibinfo{author}{\bibfnamefont{T.}~\bibnamefont{Papenbrock}},
  \bibnamefont{and}
  \bibinfo{author}{\bibfnamefont{S.}~\bibnamefont{Quaglioni}},
  \bibinfo{journal}{Phys. Rev. Lett.} \textbf{\bibinfo{volume}{126}},
  \bibinfo{pages}{182502} (\bibinfo{year}{2021}).

\bibitem[{\citenamefont{Belley et~al.}(2021)\citenamefont{Belley, Payne,
  Stroberg, Miyagi, and Holt}}]{Belley21}
\bibinfo{author}{\bibfnamefont{A.}~\bibnamefont{Belley}},
  \bibinfo{author}{\bibfnamefont{C.~G.} \bibnamefont{Payne}},
  \bibinfo{author}{\bibfnamefont{S.~R.} \bibnamefont{Stroberg}},
  \bibinfo{author}{\bibfnamefont{T.}~\bibnamefont{Miyagi}}, \bibnamefont{and}
  \bibinfo{author}{\bibfnamefont{J.~D.} \bibnamefont{Holt}},
  \bibinfo{journal}{Phys. Rev. Lett.} \textbf{\bibinfo{volume}{126}},
  \bibinfo{pages}{042502} (\bibinfo{year}{2021}).

\bibitem[{\citenamefont{Barea et~al.}(2013)\citenamefont{Barea, Kotila, and
  Iachello}}]{Barea13}
\bibinfo{author}{\bibfnamefont{J.}~\bibnamefont{Barea}},
  \bibinfo{author}{\bibfnamefont{J.}~\bibnamefont{Kotila}}, \bibnamefont{and}
  \bibinfo{author}{\bibfnamefont{F.}~\bibnamefont{Iachello}},
  \bibinfo{journal}{Phys. Rev. C} \textbf{\bibinfo{volume}{87}},
  \bibinfo{pages}{014315} (\bibinfo{year}{2013}).

\bibitem[{\citenamefont{Terasaki}(2015)}]{Terasaki15}
\bibinfo{author}{\bibfnamefont{J.}~\bibnamefont{Terasaki}},
  \bibinfo{journal}{Phys. Rev. C} \textbf{\bibinfo{volume}{91}},
  \bibinfo{pages}{034318} (\bibinfo{year}{2015}).

\bibitem[{\citenamefont{Fang et~al.}(2018)\citenamefont{Fang, Faessler, and
  \ifmmode~\check{S}\else \v{S}\fi{}imkovic}}]{Fang18}
\bibinfo{author}{\bibfnamefont{D.-L.} \bibnamefont{Fang}},
  \bibinfo{author}{\bibfnamefont{A.}~\bibnamefont{Faessler}}, \bibnamefont{and}
  \bibinfo{author}{\bibfnamefont{F.}~\bibnamefont{\ifmmode~\check{S}\else
  \v{S}\fi{}imkovic}}, \bibinfo{journal}{Phys. Rev. C}
  \textbf{\bibinfo{volume}{97}}, \bibinfo{pages}{045503}
  (\bibinfo{year}{2018}).

\bibitem[{\citenamefont{Rodr\'{\i}guez and
  Mart\'{\i}nez-Pinedo}(2010)}]{Rodriguez10}
\bibinfo{author}{\bibfnamefont{T.~R.} \bibnamefont{Rodr\'{\i}guez}}
  \bibnamefont{and}
  \bibinfo{author}{\bibfnamefont{G.}~\bibnamefont{Mart\'{\i}nez-Pinedo}},
  \bibinfo{journal}{Phys. Rev. Lett.} \textbf{\bibinfo{volume}{105}},
  \bibinfo{pages}{252503} (\bibinfo{year}{2010}).

\bibitem[{\citenamefont{Rodr{\'i}guez and Gabriel}(2013)}]{Rodriguez13}
\bibinfo{author}{\bibfnamefont{T.~R.} \bibnamefont{Rodr{\'i}guez}}
  \bibnamefont{and} \bibinfo{author}{\bibfnamefont{M.-P.}
  \bibnamefont{Gabriel}}, \bibinfo{journal}{Phys. Lett. B}
  \textbf{\bibinfo{volume}{719}}, \bibinfo{pages}{174} (\bibinfo{year}{2013}).

\bibitem[{\citenamefont{Yao et~al.}(2015)\citenamefont{Yao, Song, Hagino, Ring,
  and Meng}}]{Yao15}
\bibinfo{author}{\bibfnamefont{J.~M.} \bibnamefont{Yao}},
  \bibinfo{author}{\bibfnamefont{L.~S.} \bibnamefont{Song}},
  \bibinfo{author}{\bibfnamefont{K.}~\bibnamefont{Hagino}},
  \bibinfo{author}{\bibfnamefont{P.}~\bibnamefont{Ring}}, \bibnamefont{and}
  \bibinfo{author}{\bibfnamefont{J.}~\bibnamefont{Meng}},
  \bibinfo{journal}{Phys. Rev. C} \textbf{\bibinfo{volume}{91}},
  \bibinfo{pages}{024316} (\bibinfo{year}{2015}).

\bibitem[{\citenamefont{Song et~al.}(2017)\citenamefont{Song, Yao, Ring, and
  Meng}}]{Song17}
\bibinfo{author}{\bibfnamefont{L.~S.} \bibnamefont{Song}},
  \bibinfo{author}{\bibfnamefont{J.~M.} \bibnamefont{Yao}},
  \bibinfo{author}{\bibfnamefont{P.}~\bibnamefont{Ring}}, \bibnamefont{and}
  \bibinfo{author}{\bibfnamefont{J.}~\bibnamefont{Meng}},
  \bibinfo{journal}{Phys. Rev. C} \textbf{\bibinfo{volume}{95}},
  \bibinfo{pages}{024305} (\bibinfo{year}{2017}).

\bibitem[{\citenamefont{Jiao et~al.}(2017)\citenamefont{Jiao, Engel, and
  Holt}}]{Jiao17}
\bibinfo{author}{\bibfnamefont{C.~F.} \bibnamefont{Jiao}},
  \bibinfo{author}{\bibfnamefont{J.}~\bibnamefont{Engel}}, \bibnamefont{and}
  \bibinfo{author}{\bibfnamefont{J.~D.} \bibnamefont{Holt}},
  \bibinfo{journal}{Phys. Rev. C} \textbf{\bibinfo{volume}{96}},
  \bibinfo{pages}{054310} (\bibinfo{year}{2017}).

\bibitem[{\citenamefont{Jiao et~al.}(2018)\citenamefont{Jiao, Horoi, and
  Neacsu}}]{Jiao18}
\bibinfo{author}{\bibfnamefont{C.~F.} \bibnamefont{Jiao}},
  \bibinfo{author}{\bibfnamefont{M.}~\bibnamefont{Horoi}}, \bibnamefont{and}
  \bibinfo{author}{\bibfnamefont{A.}~\bibnamefont{Neacsu}},
  \bibinfo{journal}{Phys. Rev. C} \textbf{\bibinfo{volume}{98}},
  \bibinfo{pages}{064324} (\bibinfo{year}{2018}).

\bibitem[{\citenamefont{Sen'kov and Horoi}(2013)}]{Senkov13}
\bibinfo{author}{\bibfnamefont{R.~A.} \bibnamefont{Sen'kov}} \bibnamefont{and}
  \bibinfo{author}{\bibfnamefont{M.}~\bibnamefont{Horoi}},
  \bibinfo{journal}{Phys. Rev. C} \textbf{\bibinfo{volume}{88}},
  \bibinfo{pages}{064312} (\bibinfo{year}{2013}).

\bibitem[{\citenamefont{Holt and Engel}(2013)}]{Holt13d}
\bibinfo{author}{\bibfnamefont{J.~D.} \bibnamefont{Holt}} \bibnamefont{and}
  \bibinfo{author}{\bibfnamefont{J.}~\bibnamefont{Engel}},
  \bibinfo{journal}{Phys. Rev. C} \textbf{\bibinfo{volume}{87}},
  \bibinfo{pages}{064315} (\bibinfo{year}{2013}).

\bibitem[{\citenamefont{Sen'kov et~al.}(2014)\citenamefont{Sen'kov, Horoi, and
  Brown}}]{Senkov14}
\bibinfo{author}{\bibfnamefont{R.~A.} \bibnamefont{Sen'kov}},
  \bibinfo{author}{\bibfnamefont{M.}~\bibnamefont{Horoi}}, \bibnamefont{and}
  \bibinfo{author}{\bibfnamefont{B.~A.} \bibnamefont{Brown}},
  \bibinfo{journal}{Phys. Rev. C} \textbf{\bibinfo{volume}{89}},
  \bibinfo{pages}{054304} (\bibinfo{year}{2014}).

\bibitem[{\citenamefont{Neacsu and Horoi}(2015)}]{Neacsu15}
\bibinfo{author}{\bibfnamefont{A.}~\bibnamefont{Neacsu}} \bibnamefont{and}
  \bibinfo{author}{\bibfnamefont{M.}~\bibnamefont{Horoi}},
  \bibinfo{journal}{Phys. Rev. C} \textbf{\bibinfo{volume}{91}},
  \bibinfo{pages}{024309} (\bibinfo{year}{2015}).

\bibitem[{\citenamefont{Men{\'{e}}ndez}(2017)}]{Menendez17}
\bibinfo{author}{\bibfnamefont{J.}~\bibnamefont{Men{\'{e}}ndez}},
  \bibinfo{journal}{J. Phys. G} \textbf{\bibinfo{volume}{45}},
  \bibinfo{pages}{014003} (\bibinfo{year}{2017}).

\bibitem[{\citenamefont{Coraggio
  et~al.}(2020{\natexlab{a}})\citenamefont{Coraggio, Itaco, De~Gregorio,
  Gargano, Mancino, and Pastore}}]{Coraggio20b}
\bibinfo{author}{\bibfnamefont{L.}~\bibnamefont{Coraggio}},
  \bibinfo{author}{\bibfnamefont{N.}~\bibnamefont{Itaco}},
  \bibinfo{author}{\bibfnamefont{G.}~\bibnamefont{De~Gregorio}},
  \bibinfo{author}{\bibfnamefont{A.}~\bibnamefont{Gargano}},
  \bibinfo{author}{\bibfnamefont{R.}~\bibnamefont{Mancino}}, \bibnamefont{and}
  \bibinfo{author}{\bibfnamefont{S.}~\bibnamefont{Pastore}},
  \bibinfo{journal}{Universe} \textbf{\bibinfo{volume}{6}},
  \bibinfo{pages}{233} (\bibinfo{year}{2020}{\natexlab{a}}).

\bibitem[{\citenamefont{Wang et~al.}(2017)\citenamefont{Wang, Audi, Kondev,
  Huang, Naimi, and Xing}}]{Wang17}
\bibinfo{author}{\bibfnamefont{M.}~\bibnamefont{Wang}},
  \bibinfo{author}{\bibfnamefont{G.}~\bibnamefont{Audi}},
  \bibinfo{author}{\bibfnamefont{F.~G.} \bibnamefont{Kondev}},
  \bibinfo{author}{\bibfnamefont{W.~J.} \bibnamefont{Huang}},
  \bibinfo{author}{\bibfnamefont{S.}~\bibnamefont{Naimi}}, \bibnamefont{and}
  \bibinfo{author}{\bibfnamefont{X.}~\bibnamefont{Xing}},
  \bibinfo{journal}{Chin. Phys. C} \textbf{\bibinfo{volume}{41}},
  \bibinfo{pages}{030003} (\bibinfo{year}{2017}).

\bibitem[{\citenamefont{Alenkov et~al.}(2019)\citenamefont{Alenkov, Bae, Beyer,
  Boiko, Boonin, Buzanov, Chanthima, Cheoun, Chernyak, Choe
  et~al.}}]{Alenkov19}
\bibinfo{author}{\bibfnamefont{V.}~\bibnamefont{Alenkov}},
  \bibinfo{author}{\bibfnamefont{H.~W.} \bibnamefont{Bae}},
  \bibinfo{author}{\bibfnamefont{J.}~\bibnamefont{Beyer}},
  \bibinfo{author}{\bibfnamefont{R.~S.} \bibnamefont{Boiko}},
  \bibinfo{author}{\bibfnamefont{K.}~\bibnamefont{Boonin}},
  \bibinfo{author}{\bibfnamefont{O.}~\bibnamefont{Buzanov}},
  \bibinfo{author}{\bibfnamefont{N.}~\bibnamefont{Chanthima}},
  \bibinfo{author}{\bibfnamefont{M.~K.} \bibnamefont{Cheoun}},
  \bibinfo{author}{\bibfnamefont{D.~M.} \bibnamefont{Chernyak}},
  \bibinfo{author}{\bibfnamefont{J.~S.} \bibnamefont{Choe}},
  \bibnamefont{et~al.}, \bibinfo{journal}{Eur. Phys. J. C}
  \textbf{\bibinfo{volume}{79}}, \bibinfo{pages}{791} (\bibinfo{year}{2019}).

\bibitem[{\citenamefont{Bhang et~al.}(2012)\citenamefont{Bhang, Boiko,
  Chernyak, Choi, Choi, Danevich, Efendiev, Enss, Fleischmann, Gangapshev
  et~al.}}]{Bhang12}
\bibinfo{author}{\bibfnamefont{H.}~\bibnamefont{Bhang}},
  \bibinfo{author}{\bibfnamefont{R.~S.} \bibnamefont{Boiko}},
  \bibinfo{author}{\bibfnamefont{D.~M.} \bibnamefont{Chernyak}},
  \bibinfo{author}{\bibfnamefont{J.~H.} \bibnamefont{Choi}},
  \bibinfo{author}{\bibfnamefont{S.}~\bibnamefont{Choi}},
  \bibinfo{author}{\bibfnamefont{F.~A.} \bibnamefont{Danevich}},
  \bibinfo{author}{\bibfnamefont{K.~V.} \bibnamefont{Efendiev}},
  \bibinfo{author}{\bibfnamefont{C.}~\bibnamefont{Enss}},
  \bibinfo{author}{\bibfnamefont{A.}~\bibnamefont{Fleischmann}},
  \bibinfo{author}{\bibfnamefont{A.~M.} \bibnamefont{Gangapshev}},
  \bibnamefont{et~al.}, \bibinfo{journal}{Journal of Physics: Conference
  Series} \textbf{\bibinfo{volume}{375}}, \bibinfo{pages}{042023}
  (\bibinfo{year}{2012}).

\bibitem[{\citenamefont{Arnold et~al.}(2007)\citenamefont{Arnold, Augier,
  Baker, Barabash, Bongrand, Broudin, Brudanin, Caffrey, Egorov, Etienvre
  et~al.}}]{NEMO3}
\bibinfo{author}{\bibfnamefont{R.}~\bibnamefont{Arnold}},
  \bibinfo{author}{\bibfnamefont{C.}~\bibnamefont{Augier}},
  \bibinfo{author}{\bibfnamefont{J.}~\bibnamefont{Baker}},
  \bibinfo{author}{\bibfnamefont{A.}~\bibnamefont{Barabash}},
  \bibinfo{author}{\bibfnamefont{M.}~\bibnamefont{Bongrand}},
  \bibinfo{author}{\bibfnamefont{G.}~\bibnamefont{Broudin}},
  \bibinfo{author}{\bibfnamefont{V.}~\bibnamefont{Brudanin}},
  \bibinfo{author}{\bibfnamefont{A.}~\bibnamefont{Caffrey}},
  \bibinfo{author}{\bibfnamefont{V.}~\bibnamefont{Egorov}},
  \bibinfo{author}{\bibfnamefont{A.}~\bibnamefont{Etienvre}},
  \bibnamefont{et~al.}, \bibinfo{journal}{Nucl. Phys. A}
  \textbf{\bibinfo{volume}{781}}, \bibinfo{pages}{209 } (\bibinfo{year}{2007}).

\bibitem[{\citenamefont{Armengaud et~al.}(2020)\citenamefont{Armengaud, Augier,
  Barabash, Bellini, Benato, Beno{\^i}t, Beretta, Berg{\'e}, Billard, Borovlev
  et~al.}}]{Armengaud20a}
\bibinfo{author}{\bibfnamefont{E.}~\bibnamefont{Armengaud}},
  \bibinfo{author}{\bibfnamefont{C.}~\bibnamefont{Augier}},
  \bibinfo{author}{\bibfnamefont{A.~S.} \bibnamefont{Barabash}},
  \bibinfo{author}{\bibfnamefont{F.}~\bibnamefont{Bellini}},
  \bibinfo{author}{\bibfnamefont{G.}~\bibnamefont{Benato}},
  \bibinfo{author}{\bibfnamefont{A.}~\bibnamefont{Beno{\^i}t}},
  \bibinfo{author}{\bibfnamefont{M.}~\bibnamefont{Beretta}},
  \bibinfo{author}{\bibfnamefont{L.}~\bibnamefont{Berg{\'e}}},
  \bibinfo{author}{\bibfnamefont{J.}~\bibnamefont{Billard}},
  \bibinfo{author}{\bibfnamefont{Y.~A.} \bibnamefont{Borovlev}},
  \bibnamefont{et~al.} (\bibinfo{collaboration}{CUPID-Mo Collaboration}),
  \bibinfo{journal}{Eur. Phys. J. C} \textbf{\bibinfo{volume}{80}},
  \bibinfo{pages}{674} (\bibinfo{year}{2020}).

\bibitem[{\citenamefont{Armengaud et~al.}(2021)\citenamefont{Armengaud, Augier,
  Barabash, Bellini, Benato, Beno\^{\i}t, Beretta, Berg\'e, Billard, Borovlev
  et~al.}}]{Armengaud21}
\bibinfo{author}{\bibfnamefont{E.}~\bibnamefont{Armengaud}},
  \bibinfo{author}{\bibfnamefont{C.}~\bibnamefont{Augier}},
  \bibinfo{author}{\bibfnamefont{A.~S.} \bibnamefont{Barabash}},
  \bibinfo{author}{\bibfnamefont{F.}~\bibnamefont{Bellini}},
  \bibinfo{author}{\bibfnamefont{G.}~\bibnamefont{Benato}},
  \bibinfo{author}{\bibfnamefont{A.}~\bibnamefont{Beno\^{\i}t}},
  \bibinfo{author}{\bibfnamefont{M.}~\bibnamefont{Beretta}},
  \bibinfo{author}{\bibfnamefont{L.}~\bibnamefont{Berg\'e}},
  \bibinfo{author}{\bibfnamefont{J.}~\bibnamefont{Billard}},
  \bibinfo{author}{\bibfnamefont{Y.~A.} \bibnamefont{Borovlev}},
  \bibnamefont{et~al.} (\bibinfo{collaboration}{CUPID-Mo Collaboration}),
  \bibinfo{journal}{Phys. Rev. Lett.} \textbf{\bibinfo{volume}{126}},
  \bibinfo{pages}{181802} (\bibinfo{year}{2021}).

\bibitem[{\citenamefont{{The CUPID Interest Group}}(2019)}]{CUPID19}
\bibinfo{author}{\bibnamefont{{The CUPID Interest Group}}},
  \bibinfo{journal}{arXiv:1907.09376[physics]}  (\bibinfo{year}{2019}).

\bibitem[{\citenamefont{Cheifetz et~al.}(1970)\citenamefont{Cheifetz, Jared,
  Thompson, and Wilhelmy}}]{Cheifetz70}
\bibinfo{author}{\bibfnamefont{E.}~\bibnamefont{Cheifetz}},
  \bibinfo{author}{\bibfnamefont{R.~C.} \bibnamefont{Jared}},
  \bibinfo{author}{\bibfnamefont{S.~G.} \bibnamefont{Thompson}},
  \bibnamefont{and} \bibinfo{author}{\bibfnamefont{J.~B.}
  \bibnamefont{Wilhelmy}}, \bibinfo{journal}{Phys. Rev. Lett.}
  \textbf{\bibinfo{volume}{25}}, \bibinfo{pages}{38} (\bibinfo{year}{1970}).

\bibitem[{\citenamefont{von Brentano et~al.}(2004)\citenamefont{von Brentano,
  Werner, Casten, Scholl, McCutchan, Kr\"ucken, and Jolie}}]{vonBrentano04}
\bibinfo{author}{\bibfnamefont{P.}~\bibnamefont{von Brentano}},
  \bibinfo{author}{\bibfnamefont{V.}~\bibnamefont{Werner}},
  \bibinfo{author}{\bibfnamefont{R.~F.} \bibnamefont{Casten}},
  \bibinfo{author}{\bibfnamefont{C.}~\bibnamefont{Scholl}},
  \bibinfo{author}{\bibfnamefont{E.~A.} \bibnamefont{McCutchan}},
  \bibinfo{author}{\bibfnamefont{R.}~\bibnamefont{Kr\"ucken}},
  \bibnamefont{and} \bibinfo{author}{\bibfnamefont{J.}~\bibnamefont{Jolie}},
  \bibinfo{journal}{Phys. Rev. Lett.} \textbf{\bibinfo{volume}{93}},
  \bibinfo{pages}{152502} (\bibinfo{year}{2004}).

\bibitem[{\citenamefont{Cejnar and Jolie}(2004)}]{Cejnar04}
\bibinfo{author}{\bibfnamefont{P.}~\bibnamefont{Cejnar}} \bibnamefont{and}
  \bibinfo{author}{\bibfnamefont{J.}~\bibnamefont{Jolie}},
  \bibinfo{journal}{Phys. Rev. C} \textbf{\bibinfo{volume}{69}},
  \bibinfo{pages}{011301} (\bibinfo{year}{2004}).

\bibitem[{\citenamefont{Zhang et~al.}(2015)\citenamefont{Zhang, Bhat,
  Nazarewicz, Sheikh, and Shi}}]{Zhang15}
\bibinfo{author}{\bibfnamefont{C.~L.} \bibnamefont{Zhang}},
  \bibinfo{author}{\bibfnamefont{G.~H.} \bibnamefont{Bhat}},
  \bibinfo{author}{\bibfnamefont{W.}~\bibnamefont{Nazarewicz}},
  \bibinfo{author}{\bibfnamefont{J.~A.} \bibnamefont{Sheikh}},
  \bibnamefont{and} \bibinfo{author}{\bibfnamefont{Y.}~\bibnamefont{Shi}},
  \bibinfo{journal}{Phys. Rev. C} \textbf{\bibinfo{volume}{92}},
  \bibinfo{pages}{034307} (\bibinfo{year}{2015}).

\bibitem[{\citenamefont{Xiang et~al.}(2016)\citenamefont{Xiang, Yao, Fu, Wang,
  Li, and Long}}]{Xiang16}
\bibinfo{author}{\bibfnamefont{J.}~\bibnamefont{Xiang}},
  \bibinfo{author}{\bibfnamefont{J.~M.} \bibnamefont{Yao}},
  \bibinfo{author}{\bibfnamefont{Y.}~\bibnamefont{Fu}},
  \bibinfo{author}{\bibfnamefont{Z.~H.} \bibnamefont{Wang}},
  \bibinfo{author}{\bibfnamefont{Z.~P.} \bibnamefont{Li}}, \bibnamefont{and}
  \bibinfo{author}{\bibfnamefont{W.~H.} \bibnamefont{Long}},
  \bibinfo{journal}{Phys. Rev. C} \textbf{\bibinfo{volume}{93}},
  \bibinfo{pages}{054324} (\bibinfo{year}{2016}).

\bibitem[{\citenamefont{Abusara et~al.}(2017)\citenamefont{Abusara, Ahmad, and
  Othman}}]{Abusara17}
\bibinfo{author}{\bibfnamefont{H.}~\bibnamefont{Abusara}},
  \bibinfo{author}{\bibfnamefont{S.}~\bibnamefont{Ahmad}}, \bibnamefont{and}
  \bibinfo{author}{\bibfnamefont{S.}~\bibnamefont{Othman}},
  \bibinfo{journal}{Phys. Rev. C} \textbf{\bibinfo{volume}{95}},
  \bibinfo{pages}{054302} (\bibinfo{year}{2017}).

\bibitem[{\citenamefont{Johnstone and Towner}(1998)}]{Johnstone98}
\bibinfo{author}{\bibfnamefont{I.~P.} \bibnamefont{Johnstone}}
  \bibnamefont{and} \bibinfo{author}{\bibfnamefont{I.~S.}
  \bibnamefont{Towner}}, \bibinfo{journal}{Eur. Phys. J. A}
  \textbf{\bibinfo{volume}{3}}, \bibinfo{pages}{237} (\bibinfo{year}{1998}).

\bibitem[{\citenamefont{\"Ozen and Dean}(2006)}]{Ozen06}
\bibinfo{author}{\bibfnamefont{C.}~\bibnamefont{\"Ozen}} \bibnamefont{and}
  \bibinfo{author}{\bibfnamefont{D.~J.} \bibnamefont{Dean}},
  \bibinfo{journal}{Phys. Rev. C} \textbf{\bibinfo{volume}{73}},
  \bibinfo{pages}{014302} (\bibinfo{year}{2006}).

\bibitem[{\citenamefont{Vaquero et~al.}(2013)\citenamefont{Vaquero,
  Rodr\'{\i}guez, and Egido}}]{Vaquero13}
\bibinfo{author}{\bibfnamefont{N.~L.} \bibnamefont{Vaquero}},
  \bibinfo{author}{\bibfnamefont{T.~R.} \bibnamefont{Rodr\'{\i}guez}},
  \bibnamefont{and} \bibinfo{author}{\bibfnamefont{J.~L.} \bibnamefont{Egido}},
  \bibinfo{journal}{Phys. Rev. Lett.} \textbf{\bibinfo{volume}{111}},
  \bibinfo{pages}{142501} (\bibinfo{year}{2013}).

\bibitem[{\citenamefont{Barea et~al.}(2015)\citenamefont{Barea, Kotila, and
  Iachello}}]{Barea15}
\bibinfo{author}{\bibfnamefont{J.}~\bibnamefont{Barea}},
  \bibinfo{author}{\bibfnamefont{J.}~\bibnamefont{Kotila}}, \bibnamefont{and}
  \bibinfo{author}{\bibfnamefont{F.}~\bibnamefont{Iachello}},
  \bibinfo{journal}{Phys. Rev. C} \textbf{\bibinfo{volume}{91}},
  \bibinfo{pages}{034304} (\bibinfo{year}{2015}).

\bibitem[{\citenamefont{Tomoda}(1991)}]{Tomoda91}
\bibinfo{author}{\bibfnamefont{T.}~\bibnamefont{Tomoda}},
  \bibinfo{journal}{Rep. Prog. Phys.} \textbf{\bibinfo{volume}{54}},
  \bibinfo{pages}{53} (\bibinfo{year}{1991}).

\bibitem[{\citenamefont{Pantis et~al.}(1996)\citenamefont{Pantis,
  \ifmmode~\check{S}\else \v{S}\fi{}imkovic, Vergados, and
  Faessler}}]{Pantis96}
\bibinfo{author}{\bibfnamefont{G.}~\bibnamefont{Pantis}},
  \bibinfo{author}{\bibfnamefont{F.}~\bibnamefont{\ifmmode~\check{S}\else
  \v{S}\fi{}imkovic}}, \bibinfo{author}{\bibfnamefont{J.~D.}
  \bibnamefont{Vergados}}, \bibnamefont{and}
  \bibinfo{author}{\bibfnamefont{A.}~\bibnamefont{Faessler}},
  \bibinfo{journal}{Phys. Rev. C} \textbf{\bibinfo{volume}{53}},
  \bibinfo{pages}{695} (\bibinfo{year}{1996}).

\bibitem[{\citenamefont{Chaturvedi et~al.}(2003)\citenamefont{Chaturvedi,
  Dixit, Rath, and Raina}}]{Chaturvedi03}
\bibinfo{author}{\bibfnamefont{K.}~\bibnamefont{Chaturvedi}},
  \bibinfo{author}{\bibfnamefont{B.~M.} \bibnamefont{Dixit}},
  \bibinfo{author}{\bibfnamefont{P.~K.} \bibnamefont{Rath}}, \bibnamefont{and}
  \bibinfo{author}{\bibfnamefont{P.~K.} \bibnamefont{Raina}},
  \bibinfo{journal}{Phys. Rev. C} \textbf{\bibinfo{volume}{67}},
  \bibinfo{pages}{064317} (\bibinfo{year}{2003}).

\bibitem[{\citenamefont{\ifmmode~\check{S}\else \v{S}\fi{}imkovic
  et~al.}(2013)\citenamefont{\ifmmode~\check{S}\else \v{S}\fi{}imkovic, Rodin,
  Faessler, and Vogel}}]{Simkovic13}
\bibinfo{author}{\bibfnamefont{F.}~\bibnamefont{\ifmmode~\check{S}\else
  \v{S}\fi{}imkovic}}, \bibinfo{author}{\bibfnamefont{V.}~\bibnamefont{Rodin}},
  \bibinfo{author}{\bibfnamefont{A.}~\bibnamefont{Faessler}}, \bibnamefont{and}
  \bibinfo{author}{\bibfnamefont{P.}~\bibnamefont{Vogel}},
  \bibinfo{journal}{Phys. Rev. C} \textbf{\bibinfo{volume}{87}},
  \bibinfo{pages}{045501} (\bibinfo{year}{2013}).

\bibitem[{\citenamefont{Hyv\"arinen and Suhonen}(2015)}]{Hyvarinen15}
\bibinfo{author}{\bibfnamefont{J.}~\bibnamefont{Hyv\"arinen}} \bibnamefont{and}
  \bibinfo{author}{\bibfnamefont{J.}~\bibnamefont{Suhonen}},
  \bibinfo{journal}{Phys. Rev. C} \textbf{\bibinfo{volume}{91}},
  \bibinfo{pages}{024613} (\bibinfo{year}{2015}).

\bibitem[{\citenamefont{Coraggio et~al.}(2009)\citenamefont{Coraggio, Covello,
  Gargano, Itaco, and Kuo}}]{Coraggio09a}
\bibinfo{author}{\bibfnamefont{L.}~\bibnamefont{Coraggio}},
  \bibinfo{author}{\bibfnamefont{A.}~\bibnamefont{Covello}},
  \bibinfo{author}{\bibfnamefont{A.}~\bibnamefont{Gargano}},
  \bibinfo{author}{\bibfnamefont{N.}~\bibnamefont{Itaco}}, \bibnamefont{and}
  \bibinfo{author}{\bibfnamefont{T.~T.~S.} \bibnamefont{Kuo}},
  \bibinfo{journal}{Prog. Part. Nucl. Phys.} \textbf{\bibinfo{volume}{62}},
  \bibinfo{pages}{135} (\bibinfo{year}{2009}).

\bibitem[{\citenamefont{Machleidt}(2001)}]{Machleidt01b}
\bibinfo{author}{\bibfnamefont{R.}~\bibnamefont{Machleidt}},
  \bibinfo{journal}{Phys. Rev. C} \textbf{\bibinfo{volume}{63}},
  \bibinfo{pages}{024001} (\bibinfo{year}{2001}).

\bibitem[{\citenamefont{Bogner et~al.}(2002)\citenamefont{Bogner, Kuo,
  Coraggio, Covello, and Itaco}}]{Bogner02}
\bibinfo{author}{\bibfnamefont{S.}~\bibnamefont{Bogner}},
  \bibinfo{author}{\bibfnamefont{T.~T.~S.} \bibnamefont{Kuo}},
  \bibinfo{author}{\bibfnamefont{L.}~\bibnamefont{Coraggio}},
  \bibinfo{author}{\bibfnamefont{A.}~\bibnamefont{Covello}}, \bibnamefont{and}
  \bibinfo{author}{\bibfnamefont{N.}~\bibnamefont{Itaco}},
  \bibinfo{journal}{Phys. Rev. C} \textbf{\bibinfo{volume}{65}},
  \bibinfo{pages}{051301(R)} (\bibinfo{year}{2002}).

\bibitem[{\citenamefont{Kuo et~al.}(1995)\citenamefont{Kuo, Krmpoti\'c, Suzuki,
  and Okamoto}}]{Kuo95}
\bibinfo{author}{\bibfnamefont{T.~T.~S.} \bibnamefont{Kuo}},
  \bibinfo{author}{\bibfnamefont{F.}~\bibnamefont{Krmpoti\'c}},
  \bibinfo{author}{\bibfnamefont{K.}~\bibnamefont{Suzuki}}, \bibnamefont{and}
  \bibinfo{author}{\bibfnamefont{R.}~\bibnamefont{Okamoto}},
  \bibinfo{journal}{Nucl. Phys. A} \textbf{\bibinfo{volume}{582}},
  \bibinfo{pages}{205} (\bibinfo{year}{1995}).

\bibitem[{\citenamefont{Hjorth-Jensen et~al.}(1995)\citenamefont{Hjorth-Jensen,
  Kuo, and Osnes}}]{Hjorth95}
\bibinfo{author}{\bibfnamefont{M.}~\bibnamefont{Hjorth-Jensen}},
  \bibinfo{author}{\bibfnamefont{T.~T.~S.} \bibnamefont{Kuo}},
  \bibnamefont{and} \bibinfo{author}{\bibfnamefont{E.}~\bibnamefont{Osnes}},
  \bibinfo{journal}{Phys. Rep.} \textbf{\bibinfo{volume}{261}},
  \bibinfo{pages}{125} (\bibinfo{year}{1995}).

\bibitem[{\citenamefont{Suzuki and Okamoto}(1995)}]{Suzuki95}
\bibinfo{author}{\bibfnamefont{K.}~\bibnamefont{Suzuki}} \bibnamefont{and}
  \bibinfo{author}{\bibfnamefont{R.}~\bibnamefont{Okamoto}},
  \bibinfo{journal}{Prog. Theor. Phys.} \textbf{\bibinfo{volume}{93}},
  \bibinfo{pages}{905} (\bibinfo{year}{1995}).

\bibitem[{\citenamefont{Coraggio et~al.}(2012)\citenamefont{Coraggio, Covello,
  Gargano, Itaco, and Kuo}}]{Coraggio12a}
\bibinfo{author}{\bibfnamefont{L.}~\bibnamefont{Coraggio}},
  \bibinfo{author}{\bibfnamefont{A.}~\bibnamefont{Covello}},
  \bibinfo{author}{\bibfnamefont{A.}~\bibnamefont{Gargano}},
  \bibinfo{author}{\bibfnamefont{N.}~\bibnamefont{Itaco}}, \bibnamefont{and}
  \bibinfo{author}{\bibfnamefont{T.~T.~S.} \bibnamefont{Kuo}},
  \bibinfo{journal}{Ann. Phys. (NY)} \textbf{\bibinfo{volume}{327}},
  \bibinfo{pages}{2125} (\bibinfo{year}{2012}).

\bibitem[{\citenamefont{Ellis and Osnes}(1977)}]{Ellis77}
\bibinfo{author}{\bibfnamefont{P.~J.} \bibnamefont{Ellis}} \bibnamefont{and}
  \bibinfo{author}{\bibfnamefont{E.}~\bibnamefont{Osnes}},
  \bibinfo{journal}{Rev. Mod. Phys.} \textbf{\bibinfo{volume}{49}},
  \bibinfo{pages}{777} (\bibinfo{year}{1977}).

\bibitem[{\citenamefont{Coraggio and Itaco}(2020)}]{Coraggio20c}
\bibinfo{author}{\bibfnamefont{L.}~\bibnamefont{Coraggio}} \bibnamefont{and}
  \bibinfo{author}{\bibfnamefont{N.}~\bibnamefont{Itaco}},
  \bibinfo{journal}{Frontiers in Physics} \textbf{\bibinfo{volume}{8}},
  \bibinfo{pages}{345} (\bibinfo{year}{2020}).

\bibitem[{\citenamefont{Coraggio et~al.}(2017)\citenamefont{Coraggio,
  De~Angelis, Fukui, Gargano, and Itaco}}]{Coraggio17a}
\bibinfo{author}{\bibfnamefont{L.}~\bibnamefont{Coraggio}},
  \bibinfo{author}{\bibfnamefont{L.}~\bibnamefont{De~Angelis}},
  \bibinfo{author}{\bibfnamefont{T.}~\bibnamefont{Fukui}},
  \bibinfo{author}{\bibfnamefont{A.}~\bibnamefont{Gargano}}, \bibnamefont{and}
  \bibinfo{author}{\bibfnamefont{N.}~\bibnamefont{Itaco}},
  \bibinfo{journal}{Phys. Rev. C} \textbf{\bibinfo{volume}{95}},
  \bibinfo{pages}{064324} (\bibinfo{year}{2017}).

\bibitem[{\citenamefont{Coraggio et~al.}(2019)\citenamefont{Coraggio,
  De~Angelis, Fukui, Gargano, Itaco, and Nowacki}}]{Coraggio19a}
\bibinfo{author}{\bibfnamefont{L.}~\bibnamefont{Coraggio}},
  \bibinfo{author}{\bibfnamefont{L.}~\bibnamefont{De~Angelis}},
  \bibinfo{author}{\bibfnamefont{T.}~\bibnamefont{Fukui}},
  \bibinfo{author}{\bibfnamefont{A.}~\bibnamefont{Gargano}},
  \bibinfo{author}{\bibfnamefont{N.}~\bibnamefont{Itaco}}, \bibnamefont{and}
  \bibinfo{author}{\bibfnamefont{F.}~\bibnamefont{Nowacki}},
  \bibinfo{journal}{Phys. Rev. C} \textbf{\bibinfo{volume}{100}},
  \bibinfo{pages}{014316} (\bibinfo{year}{2019}).

\bibitem[{\citenamefont{Coraggio
  et~al.}(2020{\natexlab{b}})\citenamefont{Coraggio, Gargano, Itaco, Mancino,
  and Nowacki}}]{Coraggio20a}
\bibinfo{author}{\bibfnamefont{L.}~\bibnamefont{Coraggio}},
  \bibinfo{author}{\bibfnamefont{A.}~\bibnamefont{Gargano}},
  \bibinfo{author}{\bibfnamefont{N.}~\bibnamefont{Itaco}},
  \bibinfo{author}{\bibfnamefont{R.}~\bibnamefont{Mancino}}, \bibnamefont{and}
  \bibinfo{author}{\bibfnamefont{F.}~\bibnamefont{Nowacki}},
  \bibinfo{journal}{Phys. Rev. C} \textbf{\bibinfo{volume}{101}},
  \bibinfo{pages}{044315} (\bibinfo{year}{2020}{\natexlab{b}}).

\bibitem[{\citenamefont{Sieja et~al.}(2009)\citenamefont{Sieja, Nowacki,
  Langanke, and Mart\'{\i}nez-Pinedo}}]{Sieja09}
\bibinfo{author}{\bibfnamefont{K.}~\bibnamefont{Sieja}},
  \bibinfo{author}{\bibfnamefont{F.}~\bibnamefont{Nowacki}},
  \bibinfo{author}{\bibfnamefont{K.}~\bibnamefont{Langanke}}, \bibnamefont{and}
  \bibinfo{author}{\bibfnamefont{G.}~\bibnamefont{Mart\'{\i}nez-Pinedo}},
  \bibinfo{journal}{Phys. Rev. C} \textbf{\bibinfo{volume}{79}},
  \bibinfo{pages}{064310} (\bibinfo{year}{2009}).

\bibitem[{\citenamefont{Coraggio et~al.}(2016)\citenamefont{Coraggio, Gargano,
  and Itaco}}]{Coraggio16a}
\bibinfo{author}{\bibfnamefont{L.}~\bibnamefont{Coraggio}},
  \bibinfo{author}{\bibfnamefont{A.}~\bibnamefont{Gargano}}, \bibnamefont{and}
  \bibinfo{author}{\bibfnamefont{N.}~\bibnamefont{Itaco}},
  \bibinfo{journal}{Phys. Rev. C} \textbf{\bibinfo{volume}{93}},
  \bibinfo{pages}{064328} (\bibinfo{year}{2016}).

\bibitem[{\citenamefont{Coraggio et~al.}(2015)\citenamefont{Coraggio, Covello,
  Gargano, Itaco, and Kuo}}]{Coraggio15a}
\bibinfo{author}{\bibfnamefont{L.}~\bibnamefont{Coraggio}},
  \bibinfo{author}{\bibfnamefont{A.}~\bibnamefont{Covello}},
  \bibinfo{author}{\bibfnamefont{A.}~\bibnamefont{Gargano}},
  \bibinfo{author}{\bibfnamefont{N.}~\bibnamefont{Itaco}}, \bibnamefont{and}
  \bibinfo{author}{\bibfnamefont{T.~T.~S.} \bibnamefont{Kuo}},
  \bibinfo{journal}{Phys. Rev. C} \textbf{\bibinfo{volume}{91}},
  \bibinfo{pages}{041301} (\bibinfo{year}{2015}).

\bibitem[{\citenamefont{Coraggio
  et~al.}(2020{\natexlab{c}})\citenamefont{Coraggio, Itaco, and
  Mancino}}]{Coraggio20d}
\bibinfo{author}{\bibfnamefont{L.}~\bibnamefont{Coraggio}},
  \bibinfo{author}{\bibfnamefont{N.}~\bibnamefont{Itaco}}, \bibnamefont{and}
  \bibinfo{author}{\bibfnamefont{R.}~\bibnamefont{Mancino}},
  \bibinfo{journal}{J. Phys. Conf. Ser.} \textbf{\bibinfo{volume}{1643}},
  \bibinfo{pages}{012124} (\bibinfo{year}{2020}{\natexlab{c}}).

\bibitem[{\citenamefont{Kuo and Osnes}(1990)}]{Kuo90}
\bibinfo{author}{\bibfnamefont{T.~T.~S.} \bibnamefont{Kuo}} \bibnamefont{and}
  \bibinfo{author}{\bibfnamefont{E.}~\bibnamefont{Osnes}},
  \emph{\bibinfo{title}{Lecture Notes in Physics}}, vol. \bibinfo{volume}{364}
  (\bibinfo{publisher}{Springer-Verlag, Berlin}, \bibinfo{year}{1990}).

\bibitem[{\citenamefont{Krenciglowa and Kuo}(1974)}]{Krenciglowa74}
\bibinfo{author}{\bibfnamefont{E.~M.} \bibnamefont{Krenciglowa}}
  \bibnamefont{and} \bibinfo{author}{\bibfnamefont{T.~T.~S.}
  \bibnamefont{Kuo}}, \bibinfo{journal}{Nucl. Phys. A}
  \textbf{\bibinfo{volume}{235}}, \bibinfo{pages}{171} (\bibinfo{year}{1974}).

\bibitem[{\citenamefont{Suzuki and Lee}(1980)}]{Suzuki80}
\bibinfo{author}{\bibfnamefont{K.}~\bibnamefont{Suzuki}} \bibnamefont{and}
  \bibinfo{author}{\bibfnamefont{S.~Y.} \bibnamefont{Lee}},
  \bibinfo{journal}{Prog. Theor. Phys.} \textbf{\bibinfo{volume}{64}},
  \bibinfo{pages}{2091} (\bibinfo{year}{1980}).

\bibitem[{\citenamefont{Suzuki et~al.}(2011)\citenamefont{Suzuki, Okamoto,
  Kumagai, and Fujii}}]{Suzuki11}
\bibinfo{author}{\bibfnamefont{K.}~\bibnamefont{Suzuki}},
  \bibinfo{author}{\bibfnamefont{R.}~\bibnamefont{Okamoto}},
  \bibinfo{author}{\bibfnamefont{H.}~\bibnamefont{Kumagai}}, \bibnamefont{and}
  \bibinfo{author}{\bibfnamefont{S.}~\bibnamefont{Fujii}},
  \bibinfo{journal}{Phys. Rev. C} \textbf{\bibinfo{volume}{83}},
  \bibinfo{pages}{024304} (\bibinfo{year}{2011}).

\bibitem[{\citenamefont{Caurier
  et~al.}(2005{\natexlab{a}})\citenamefont{Caurier, Mart\'{\i}nez-Pinedo,
  Nowacki, Poves, and Zuker}}]{ANTOINE}
\bibinfo{author}{\bibfnamefont{E.}~\bibnamefont{Caurier}},
  \bibinfo{author}{\bibfnamefont{G.}~\bibnamefont{Mart\'{\i}nez-Pinedo}},
  \bibinfo{author}{\bibfnamefont{F.}~\bibnamefont{Nowacki}},
  \bibinfo{author}{\bibfnamefont{A.}~\bibnamefont{Poves}}, \bibnamefont{and}
  \bibinfo{author}{\bibfnamefont{A.~P.} \bibnamefont{Zuker}},
  \bibinfo{journal}{Rev. Mod. Phys.} \textbf{\bibinfo{volume}{77}},
  \bibinfo{pages}{427} (\bibinfo{year}{2005}{\natexlab{a}}).

\bibitem[{\citenamefont{Coraggio
  et~al.}(2020{\natexlab{d}})\citenamefont{Coraggio, De~Gregorio, Gargano,
  Itaco, Fukui, Ma, and Xu}}]{Coraggio20e}
\bibinfo{author}{\bibfnamefont{L.}~\bibnamefont{Coraggio}},
  \bibinfo{author}{\bibfnamefont{G.}~\bibnamefont{De~Gregorio}},
  \bibinfo{author}{\bibfnamefont{A.}~\bibnamefont{Gargano}},
  \bibinfo{author}{\bibfnamefont{N.}~\bibnamefont{Itaco}},
  \bibinfo{author}{\bibfnamefont{T.}~\bibnamefont{Fukui}},
  \bibinfo{author}{\bibfnamefont{Y.~Z.} \bibnamefont{Ma}}, \bibnamefont{and}
  \bibinfo{author}{\bibfnamefont{F.~R.} \bibnamefont{Xu}},
  \bibinfo{journal}{Phys. Rev. C} \textbf{\bibinfo{volume}{102}},
  \bibinfo{pages}{054326} (\bibinfo{year}{2020}{\natexlab{d}}).

\bibitem[{sup()}]{supplemental2021}
\bibinfo{note}{See Supplemental Material at [URL will be inserted by publisher]
  for the list of two-body matrix elements of the shell-model Hamiltonian
  $H_{\rm eff}$, derived for 14 and 8 valence protons and neutrons,
  respectively, namely for $^{100}$Mo.}

\bibitem[{\citenamefont{Mavromatis et~al.}(1966)\citenamefont{Mavromatis,
  Zamick, and Brown}}]{Mavromatis66}
\bibinfo{author}{\bibfnamefont{H.~A.} \bibnamefont{Mavromatis}},
  \bibinfo{author}{\bibfnamefont{L.}~\bibnamefont{Zamick}}, \bibnamefont{and}
  \bibinfo{author}{\bibfnamefont{G.~E.} \bibnamefont{Brown}},
  \bibinfo{journal}{Nucl. Phys. A} \textbf{\bibinfo{volume}{80}},
  \bibinfo{pages}{545} (\bibinfo{year}{1966}).

\bibitem[{\citenamefont{Mavromatis and Zamick}(1967)}]{Mavromatis67}
\bibinfo{author}{\bibfnamefont{H.~A.} \bibnamefont{Mavromatis}}
  \bibnamefont{and} \bibinfo{author}{\bibfnamefont{L.}~\bibnamefont{Zamick}},
  \bibinfo{journal}{Nucl. Phys. A} \textbf{\bibinfo{volume}{104}},
  \bibinfo{pages}{17} (\bibinfo{year}{1967}).

\bibitem[{\citenamefont{Federman and Zamick}(1969)}]{Federman69}
\bibinfo{author}{\bibfnamefont{P.}~\bibnamefont{Federman}} \bibnamefont{and}
  \bibinfo{author}{\bibfnamefont{L.}~\bibnamefont{Zamick}},
  \bibinfo{journal}{Phys. Rev.} \textbf{\bibinfo{volume}{177}},
  \bibinfo{pages}{1534} (\bibinfo{year}{1969}).

\bibitem[{\citenamefont{Towner and Khanna}(1983)}]{Towner83}
\bibinfo{author}{\bibfnamefont{I.~S.} \bibnamefont{Towner}} \bibnamefont{and}
  \bibinfo{author}{\bibfnamefont{K.~F.~C.} \bibnamefont{Khanna}},
  \bibinfo{journal}{Nucl. Phys. A} \textbf{\bibinfo{volume}{399}},
  \bibinfo{pages}{334} (\bibinfo{year}{1983}).

\bibitem[{\citenamefont{Towner}(1987)}]{Towner87}
\bibinfo{author}{\bibfnamefont{I.~S.} \bibnamefont{Towner}},
  \bibinfo{journal}{Phys. Rep.} \textbf{\bibinfo{volume}{155}},
  \bibinfo{pages}{263} (\bibinfo{year}{1987}).

\bibitem[{\citenamefont{Coraggio et~al.}(2018)\citenamefont{Coraggio,
  De~Angelis, Fukui, Gargano, and Itaco}}]{Coraggio18}
\bibinfo{author}{\bibfnamefont{L.}~\bibnamefont{Coraggio}},
  \bibinfo{author}{\bibfnamefont{L.}~\bibnamefont{De~Angelis}},
  \bibinfo{author}{\bibfnamefont{T.}~\bibnamefont{Fukui}},
  \bibinfo{author}{\bibfnamefont{A.}~\bibnamefont{Gargano}}, \bibnamefont{and}
  \bibinfo{author}{\bibfnamefont{N.}~\bibnamefont{Itaco}}, \bibinfo{journal}{J.
  Phys. Conf. Ser.} \textbf{\bibinfo{volume}{1056}}, \bibinfo{pages}{012012}
  (\bibinfo{year}{2018}).

\bibitem[{\citenamefont{Haxton and Stephenson~Jr.}(1984)}]{Haxton84}
\bibinfo{author}{\bibfnamefont{W.~C.} \bibnamefont{Haxton}} \bibnamefont{and}
  \bibinfo{author}{\bibfnamefont{G.~J.} \bibnamefont{Stephenson~Jr.}},
  \bibinfo{journal}{Prog. Part. Nucl. Phys.} \textbf{\bibinfo{volume}{12}},
  \bibinfo{pages}{409} (\bibinfo{year}{1984}).

\bibitem[{\citenamefont{Elliott and Petr}(2002)}]{Elliott02}
\bibinfo{author}{\bibfnamefont{S.~R.} \bibnamefont{Elliott}} \bibnamefont{and}
  \bibinfo{author}{\bibfnamefont{V.}~\bibnamefont{Petr}},
  \bibinfo{journal}{Annu. Rev. Nucl. Part. Sci.} \textbf{\bibinfo{volume}{52}},
  \bibinfo{pages}{115} (\bibinfo{year}{2002}).

\bibitem[{\citenamefont{Caurier
  et~al.}(2005{\natexlab{b}})\citenamefont{Caurier, Mart\'{\i}nez-Pinedo,
  Nowacki, Poves, and Zuker}}]{Caurier05}
\bibinfo{author}{\bibfnamefont{E.}~\bibnamefont{Caurier}},
  \bibinfo{author}{\bibfnamefont{G.}~\bibnamefont{Mart\'{\i}nez-Pinedo}},
  \bibinfo{author}{\bibfnamefont{F.}~\bibnamefont{Nowacki}},
  \bibinfo{author}{\bibfnamefont{A.}~\bibnamefont{Poves}}, \bibnamefont{and}
  \bibinfo{author}{\bibfnamefont{A.~P.} \bibnamefont{Zuker}},
  \bibinfo{journal}{Rev. Mod. Phys.} \textbf{\bibinfo{volume}{77}},
  \bibinfo{pages}{427} (\bibinfo{year}{2005}{\natexlab{b}}).

\bibitem[{\citenamefont{Engel and Men{\'e}ndez}(2017)}]{Engel17}
\bibinfo{author}{\bibfnamefont{J.}~\bibnamefont{Engel}} \bibnamefont{and}
  \bibinfo{author}{\bibfnamefont{J.}~\bibnamefont{Men{\'e}ndez}},
  \bibinfo{journal}{Rep. Prog. Phys.} \textbf{\bibinfo{volume}{80}},
  \bibinfo{pages}{046301} (\bibinfo{year}{2017}).

\bibitem[{\citenamefont{\ifmmode~\check{S}\else \v{S}\fi{}imkovic
  et~al.}(2008)\citenamefont{\ifmmode~\check{S}\else \v{S}\fi{}imkovic,
  Faessler, Rodin, Vogel, and Engel}}]{Simkovic08}
\bibinfo{author}{\bibfnamefont{F.}~\bibnamefont{\ifmmode~\check{S}\else
  \v{S}\fi{}imkovic}},
  \bibinfo{author}{\bibfnamefont{A.}~\bibnamefont{Faessler}},
  \bibinfo{author}{\bibfnamefont{V.}~\bibnamefont{Rodin}},
  \bibinfo{author}{\bibfnamefont{P.}~\bibnamefont{Vogel}}, \bibnamefont{and}
  \bibinfo{author}{\bibfnamefont{J.}~\bibnamefont{Engel}},
  \bibinfo{journal}{Phys. Rev. C} \textbf{\bibinfo{volume}{77}},
  \bibinfo{pages}{045503} (\bibinfo{year}{2008}).

\bibitem[{\citenamefont{Bethe}(1971)}]{Bethe71}
\bibinfo{author}{\bibfnamefont{H.~A.} \bibnamefont{Bethe}},
  \bibinfo{journal}{Annu. Rev. Nucl. Sci.} \textbf{\bibinfo{volume}{21}},
  \bibinfo{pages}{93} (\bibinfo{year}{1971}).

\bibitem[{\citenamefont{Kortelainen et~al.}(2007)\citenamefont{Kortelainen,
  Civitarese, Suhonen, and Toivanen}}]{Kortelainen07}
\bibinfo{author}{\bibfnamefont{M.}~\bibnamefont{Kortelainen}},
  \bibinfo{author}{\bibfnamefont{O.}~\bibnamefont{Civitarese}},
  \bibinfo{author}{\bibfnamefont{J.}~\bibnamefont{Suhonen}}, \bibnamefont{and}
  \bibinfo{author}{\bibfnamefont{J.}~\bibnamefont{Toivanen}},
  \bibinfo{journal}{Phys. Lett. B} \textbf{\bibinfo{volume}{647}},
  \bibinfo{pages}{128} (\bibinfo{year}{2007}).

\bibitem[{\citenamefont{Men\'endez et~al.}(2009)\citenamefont{Men\'endez,
  Poves, Caurier, and Nowacki}}]{Menendez09b}
\bibinfo{author}{\bibfnamefont{J.}~\bibnamefont{Men\'endez}},
  \bibinfo{author}{\bibfnamefont{A.}~\bibnamefont{Poves}},
  \bibinfo{author}{\bibfnamefont{E.}~\bibnamefont{Caurier}}, \bibnamefont{and}
  \bibinfo{author}{\bibfnamefont{F.}~\bibnamefont{Nowacki}},
  \bibinfo{journal}{Nucl. Phys. A} \textbf{\bibinfo{volume}{818}},
  \bibinfo{pages}{139} (\bibinfo{year}{2009}).

\bibitem[{ens()}]{ensdf}
\bibinfo{note}{Data extracted using the NNDC On-line Data Service from the
  ENSDF database, file revised as of March 13, 2021.},
  \urlprefix\url{https://www.nndc.bnl.gov/ensdf}.

\bibitem[{\citenamefont{Barabash}(2020)}]{Barabash20}
\bibinfo{author}{\bibfnamefont{A.}~\bibnamefont{Barabash}},
  \bibinfo{journal}{Universe} \textbf{\bibinfo{volume}{6}},
  \bibinfo{pages}{159} (\bibinfo{year}{2020}).

\bibitem[{\citenamefont{Thies et~al.}(2012)\citenamefont{Thies, Adachi, Dozono,
  Ejiri, Frekers, Fujita, Fujita, Fujiwara, Grewe, Hatanaka et~al.}}]{Thies12c}
\bibinfo{author}{\bibfnamefont{J.~H.} \bibnamefont{Thies}},
  \bibinfo{author}{\bibfnamefont{T.}~\bibnamefont{Adachi}},
  \bibinfo{author}{\bibfnamefont{M.}~\bibnamefont{Dozono}},
  \bibinfo{author}{\bibfnamefont{H.}~\bibnamefont{Ejiri}},
  \bibinfo{author}{\bibfnamefont{D.}~\bibnamefont{Frekers}},
  \bibinfo{author}{\bibfnamefont{H.}~\bibnamefont{Fujita}},
  \bibinfo{author}{\bibfnamefont{Y.}~\bibnamefont{Fujita}},
  \bibinfo{author}{\bibfnamefont{M.}~\bibnamefont{Fujiwara}},
  \bibinfo{author}{\bibfnamefont{E.-W.} \bibnamefont{Grewe}},
  \bibinfo{author}{\bibfnamefont{K.}~\bibnamefont{Hatanaka}},
  \bibnamefont{et~al.}, \bibinfo{journal}{Phys. Rev. C}
  \textbf{\bibinfo{volume}{86}}, \bibinfo{pages}{044309}
  (\bibinfo{year}{2012}).

\bibitem[{\citenamefont{Puppe et~al.}(2012)\citenamefont{Puppe, Lennarz,
  Adachi, Akimune, Ejiri, Frekers, Fujita, Fujita, Fujiwara,
  Ganio\ifmmode~\breve{g}\else \u{g}\fi{}lu et~al.}}]{Puppe12}
\bibinfo{author}{\bibfnamefont{P.}~\bibnamefont{Puppe}},
  \bibinfo{author}{\bibfnamefont{A.}~\bibnamefont{Lennarz}},
  \bibinfo{author}{\bibfnamefont{T.}~\bibnamefont{Adachi}},
  \bibinfo{author}{\bibfnamefont{H.}~\bibnamefont{Akimune}},
  \bibinfo{author}{\bibfnamefont{H.}~\bibnamefont{Ejiri}},
  \bibinfo{author}{\bibfnamefont{D.}~\bibnamefont{Frekers}},
  \bibinfo{author}{\bibfnamefont{H.}~\bibnamefont{Fujita}},
  \bibinfo{author}{\bibfnamefont{Y.}~\bibnamefont{Fujita}},
  \bibinfo{author}{\bibfnamefont{M.}~\bibnamefont{Fujiwara}},
  \bibinfo{author}{\bibfnamefont{E.}~\bibnamefont{Ganio\ifmmode~\breve{g}\else
  \u{g}\fi{}lu}}, \bibnamefont{et~al.}, \bibinfo{journal}{Phys. Rev. C}
  \textbf{\bibinfo{volume}{86}}, \bibinfo{pages}{044603}
  (\bibinfo{year}{2012}).

\bibitem[{\citenamefont{Frekers et~al.}(2013)\citenamefont{Frekers, Puppe,
  Thies, and Ejiri}}]{Frekers13}
\bibinfo{author}{\bibfnamefont{D.}~\bibnamefont{Frekers}},
  \bibinfo{author}{\bibfnamefont{P.}~\bibnamefont{Puppe}},
  \bibinfo{author}{\bibfnamefont{J.~H.} \bibnamefont{Thies}}, \bibnamefont{and}
  \bibinfo{author}{\bibfnamefont{H.}~\bibnamefont{Ejiri}},
  \bibinfo{journal}{Nucl. Phys. A} \textbf{\bibinfo{volume}{916}},
  \bibinfo{pages}{219} (\bibinfo{year}{2013}).

\bibitem[{\citenamefont{Baker and Gammel}(1970)}]{Baker70}
\bibinfo{author}{\bibfnamefont{G.~A.} \bibnamefont{Baker}} \bibnamefont{and}
  \bibinfo{author}{\bibfnamefont{J.~L.} \bibnamefont{Gammel}},
  \emph{\bibinfo{title}{The Pad{\'e} Approximant in Theoretical Physics}},
  vol.~\bibinfo{volume}{71} of \emph{\bibinfo{series}{Mathematics in Science
  and Engineering}} (\bibinfo{publisher}{Academic Press, New York},
  \bibinfo{year}{1970}).

\bibitem[{\citenamefont{Suhonen}(2017{\natexlab{a}})}]{Suhonen17a}
\bibinfo{author}{\bibfnamefont{J.}~\bibnamefont{Suhonen}},
  \bibinfo{journal}{Phys. Rev. C} \textbf{\bibinfo{volume}{96}},
  \bibinfo{pages}{055501} (\bibinfo{year}{2017}{\natexlab{a}}).

\bibitem[{\citenamefont{Suhonen}(2017{\natexlab{b}})}]{Suhonen17b}
\bibinfo{author}{\bibfnamefont{J.~T.} \bibnamefont{Suhonen}},
  \bibinfo{journal}{Frontiers in Physics} \textbf{\bibinfo{volume}{5}},
  \bibinfo{pages}{55} (\bibinfo{year}{2017}{\natexlab{b}}).

\bibitem[{\citenamefont{Fukui et~al.}(2018)\citenamefont{Fukui, De~Angelis, Ma,
  Coraggio, Gargano, Itaco, and Xu}}]{Fukui18}
\bibinfo{author}{\bibfnamefont{T.}~\bibnamefont{Fukui}},
  \bibinfo{author}{\bibfnamefont{L.}~\bibnamefont{De~Angelis}},
  \bibinfo{author}{\bibfnamefont{Y.~Z.} \bibnamefont{Ma}},
  \bibinfo{author}{\bibfnamefont{L.}~\bibnamefont{Coraggio}},
  \bibinfo{author}{\bibfnamefont{A.}~\bibnamefont{Gargano}},
  \bibinfo{author}{\bibfnamefont{N.}~\bibnamefont{Itaco}}, \bibnamefont{and}
  \bibinfo{author}{\bibfnamefont{F.~R.} \bibnamefont{Xu}},
  \bibinfo{journal}{Phys. Rev. C} \textbf{\bibinfo{volume}{98}},
  \bibinfo{pages}{044305} (\bibinfo{year}{2018}).

\bibitem[{\citenamefont{Ma et~al.}(2019)\citenamefont{Ma, Coraggio, De~Angelis,
  Fukui, Gargano, Itaco, and Xu}}]{Ma19}
\bibinfo{author}{\bibfnamefont{Y.~Z.} \bibnamefont{Ma}},
  \bibinfo{author}{\bibfnamefont{L.}~\bibnamefont{Coraggio}},
  \bibinfo{author}{\bibfnamefont{L.}~\bibnamefont{De~Angelis}},
  \bibinfo{author}{\bibfnamefont{T.}~\bibnamefont{Fukui}},
  \bibinfo{author}{\bibfnamefont{A.}~\bibnamefont{Gargano}},
  \bibinfo{author}{\bibfnamefont{N.}~\bibnamefont{Itaco}}, \bibnamefont{and}
  \bibinfo{author}{\bibfnamefont{F.~R.} \bibnamefont{Xu}},
  \bibinfo{journal}{Phys. Rev. C} \textbf{\bibinfo{volume}{100}},
  \bibinfo{pages}{034324} (\bibinfo{year}{2019}).

\end{thebibliography}

\end{document}